\documentclass[fleqn,usenatbib]{mnras}
\usepackage[T1]{fontenc}
\usepackage{ae,aecompl}
\usepackage{graphicx}	
\usepackage{amsmath}	
\usepackage{amssymb}	
\usepackage{graphicx}
\usepackage{subfigure}
\usepackage{multirow}
\usepackage{multicol}
\usepackage{array,tabularx}
\usepackage{newtxtext,newtxmath}

\newcommand{\bvec}[1]{\mathbf{#1}}

\newcommand{\sa}[1]{{\color{black}#1}}

\newcommand{\EC}[1]{\color{black}#1}

\title[AGN Variability Through RMHD Simulations]{Numerical Analysis of Long-term Variability of AGN Jets through RMHD Simulations}

\author[S. Acharya et al.]{
Sriyasriti Acharya\thanks{E-mail: sriya.acharya@gmail.com},
Nikhil S Borse,
Bhargav Vaidya
\\
Discipline of Astronomy Astrophysics and Space Engineering, Indian Institute of Technology Indore, Khandwa Road, Simrol, Indore 453552, India
}

\date{Accepted XXX. Received YYY; in original form ZZZ}

\pubyear{2021}

\begin{document}
\label{firstpage}
\pagerange{\pageref{firstpage}--\pageref{lastpage}}
\maketitle

\begin{abstract}

Relativistic AGN jets exhibit multi-timescale variability and a broadband non-thermal spectrum extending from radio to gamma-rays. These highly magnetized jets are prone to undergo several Magneto-hydrodynamic (MHD) instabilities during their propagation in space and could trigger jet radiation and particle acceleration. This work aims to study the implications of relativistic kink mode instability on the observed long-term variability in the context of the twisting in-homogeneous jet model. To achieve this, we investigate the physical configurations preferable for forming kink mode instability by performing high-resolution 3D relativistic MHD simulations of a portion of highly magnetized jets.
In particular, we perform simulations of cylindrical plasma column with Lorentz factor $\geq 5$ and study the effects of magnetization values and axial wave-numbers with decreasing pitch on the onset and growth of kink instability. 
We have confirmed the impact of axial wave-number on the dynamics of the plasma column including the growth of the instability. 
In this work, we have further investigated the connection between the dynamics of the plasma column with its time-varying emission features.
From our analysis, we find a correlated trend between the growth rate of kink mode instability and the flux variability obtained from the simulated light curve.

\end{abstract}


\section{Introduction}

Relativistic magnetized jets are common features seen among the Active Galactic Nuclei (AGNs) \citep{urry_1995, Blandford_review_2019, Hardcastle_2020}, which lie in a direction perpendicular to the plane of the underlying accretion disk. 
Typical blazar jets that make a very small angle with respect to our line of sight emit non-thermal emission covering the whole gamut of the electromagnetic spectrum starting from radio to $\gamma$-rays along with being characterized by multi-timescale flux variability.  
It exhibits a high degree of linear polarization, revealing the presence of ordered magnetic fields that play a significant role in the jet formation, collimation, and acceleration \citep{Homan_2015, Pushkarev_2017, Blinov_Robopol_2018, Bottcher_2019}.
These magnetically driven jets dissipate energy and remain stable at parsec scales, although they suffer from a number of instabilities during their propagation in space. Current driven (CD) magneto-hydrodynamical (MHD) instabilities \citep{Appl_2000} are believed to be one of the plausible mechanisms for dissipation of magnetic energy that could possibly trigger jet radiation and particle acceleration.

Previously, several numerical simulations have been performed to understand the effect of the instabilities on the jet structure and evolution. 
In the series of papers by \cite{Mizuno_2009, Mizuno_2011, Mizuno_2012}, the influence of the initial configuration, the velocity shear, and the jet rotation has been investigated in the presence of kink instability. 
\EC{Shear driven Kelvin-Helmholtz instabilities (KHI) have also been studied in the context of stability and collimation of extra-galactic jets, both through linear and non-linear analysis \citep[see, e.g.,][]{Bodo_1989, Birkinshaw:1991, Hardee_1992, Malagoli1996, Ryu2000, Kersale2000, Perucho_2004, Mizuno_2007, Perucho_2010}.}
\cite{Tchekhovskoy_2016} have carried out global 3D MHD simulations of relativistic jets and showed that kink instability could effectively govern the morphological dichotomy of Fanaroff–Riley (FR) jets. However, understanding the impact of such MHD instabilities on the observed phenomena is still an open field.
\cite{Zhang_2017} and \cite{Bodo_2020} have performed polarization dependent radiation study of kink instability through RMHD simulations. Recently, \cite{Dong_2020} have found quasi-periodic nature in the light curve from the strongest kink region, with anti-correlated polarization signatures. 
\cite{Borse_2020} studied the effect of KHI on kpc scales, driven by the velocity shear between the jet and the ambient medium, that leads to energy dissipation and subsequently to the formation of a non-thermal electron population and the observed high energy emission.
The MHD instabilities can lead to the formation of shocks, and could trigger the onset of magnetic reconnection. These micro/macro physical processes may provide a possible explanation of observed flares and short term flux variations \citep{Marscher_Gear_1985, Gupta_2019, Giannios_2010, Ghisellini_2008_JetinJet, Giannios_2013, Striani_2016, Bottcher_2019}. \cite{Chandra_Singh_2016} also found that kink instability could initiate fast magnetic reconnection, providing an efficient way of particle acceleration in AGN jets and gamma-ray-burst jets.

Further along with short term variability features, understanding the physical processes leading to long-term variability in AGN jets and particularly in blazars is also essential. 
A curved or helical jet has given a possible explanation for the observed long-term flux variation with the help of geometrical models \citep{Villata_1998, Rieger_2004, Raiteri_2017}. Such a helical jet can be produced due to a binary black hole system (BBHS) or due to MHD instabilities. \cite{Ostorero_2004} also interpreted the long-term variability of AO $0235+16$ in terms of a helical, in-homogeneous, non-thermally emitting jet. \cite{Raiteri_2017} used an in-homogeneous twisted jet model to explain the spectral and timing properties of CTA 102. The long-term variability along with optical outburst of CTA 102 are well explained with the helical jet model through the Doppler boosting effects by varying the viewing angle. 
This motivates us to simulate relativistic magnetic kink instability and study its impact on the variability and emission features of AGN jets. As a consequence of the kink instability, the bending takes the structure of a helical or a curved jet.

The main goal is to bridge the gaps in our understanding of the underlying physical processes responsible for the observed long-term variation by doing numerical simulations. In this work, we focus on understanding the impact of MHD instabilities on the dynamical evolution and its consequences on the emission signatures. Since the development of an instability is a local feature, we simulate a particular section of a jet that could undergo instabilities. 
\cite{Zhang_2017} have studied the polarization-dependent radiation modeling of the kink instability in the blazar emission region and found that the flares or fluctuations observed in the polarization angle of the emitting region of blazar jets may have a kink origin. Recent studies have demonstrated the role of kink in generating current sheets and effect of particle acceleration due to re-connection on emission and polarisation through simulations of plasma column \citep{Bodo_2020, Kadowaki_2021, Dalpino_2021}. \EC{Recently, \cite{Kadowaki_2021} has performed 3D special relativistic MHD simulations and suggested that high energy emission variability could be originated through fast magnetic reconnection driven by kink instability.}
It should be noted that, in this study, we have presented an ideal scenario of the ``helical jet model" focusing on a small section of the jet.
\cite{Dong_2020} have simulated the jet from its central engine as it propagates through the surrounding medium with a Lorentz factor $\approx$ 2. Whereas we have simulated a column that would represent a particular section of an AGN jet by adopting a much higher Lorentz factor $\approx$ 5 and 10, which are the typical values for Blazar jets \citep{Jorstad_2005}.

The paper is structured as follows.
In section \ref{model_kink}, we describe the physical assumptions standing out for the numerical setup in order to model relativistic kink instability and its emission properties. In section \ref{results}, we present our results obtained from the parameter study along with the jet dynamical and morphological evolution. In section \ref{helical_jet}, we explain the impact of kink on the emission and variability and also discuss the results obtained from different statistical tests. Finally, in sections \ref{discussion} and \ref{summary}, we describe and summarize our current findings and discuss about our subsequent works.

\section{Modeling Relativistic Kink Instability}\label{model_kink}

\subsection{Current driven mode} \label{sec:CD_mode}

Current-driven (CD) kink mode falls under intrinsic MHD instabilities that are mainly related to the structural arrangement of magnetic field lines \citep{Lyubarskii_1999, Appl_2000, Anjiri_2014}. The magnetic pitch parameter, defined as the ratio of poloidal and toroidal magnetic field strength, plays an important role in triggering CD mode instability. In the kink instability, 
the toroidal magnetic field lines are compressed on the inner side of the plasma column and that increases the magnetic pressure. When the magnetic pressure becomes larger than the net magnetic tension, it leads to the helical displacement of the plasma column and sometimes, it may completely disrupt the system depending on its growth rate.

Magnetic kink instabilities tend to grow on the surfaces satisfying the condition $\bvec{k}.\bvec{B} = 0$, where $\bvec{k}$ is the propagation wavevector of the perturbation with magnitude $|\bvec{k}| = \frac{2\pi \rm n}{\rm L}$, $\bvec{B}$ is the magnetic field vector, $\rm L$ is the characteristic length and $\rm n$ is the wavenumber.
On these surfaces, the stabilizing effect of the magnetic tension is absent and these are often known as resonant surfaces \citep{Bodo_2013}. 
We further define resonant surfaces in cylindrical geometry \citep{Bodo_2013} as surfaces where the condition 
$$ kP + m = 0 $$ is satisfied, with $\rm m$ as the azimuthal wave-number, $\rm k$ being the wave-vector of the instability and $\rm P$ is the magnetic pitch.
The wavelengths at which the kink instability ($m=\pm 1$) can grow is $\rm k = \rm \mp m/P$. From linear analysis, irrespective of the pitch profile, the wave-number that can fit into a simulation box to attain the maximum growth in the rest frame of kink is $\rm k_{\rm max} \approx 0.745 \times 1/\rm P_{0}$ with a growth rate of $\eta_{{\rm gr}_{\rm max}} \approx 0.133 v_{\rm A}/P_{0}$,  where $\rm P_{0}$ is the pitch value at the axis and $v_{\rm A}$ is the Alfv\'en speed. 
However, in the lab frame the growth rate is expected to reduce with $\eta_{\rm gr} \propto 1/\gamma_{\rm k}$, where $\gamma_{\rm k}$ is the Lorentz factor of the moving kink \citep{Appl_2000,Bromberg_2019}.

CD kink instability is considered to be one of the possible mechanisms that internally dissipates magnetic energy associated with the Poynting flux.
Let us consider a cylindrical plasma column in a Cartesian box ($X, Y, Z$), such that the axis is along $\hat{z}$. We can study the growth of the kink instability from the perspective of energetics following the temporal evolution of the volume-averaged kinetic energy ($\rm E_{\rm kin,\rm xy}$) and magnetic energy ($\rm E_{\rm mag,\rm xy}$), that are defined in the X-Y plane as:
\begin{equation}
    E_{\rm kin,\rm xy} = \frac{1}{V_{\rm j}} \int \rho\,\frac{(v_{\rm x}^{2} + v_{\rm y}^{2})}{2} \,dV,
    \label{eq:kin_xy}
\end{equation}
and
\begin{equation}
    E_{\rm mag,\rm xy} = \frac{1}{V_{\rm j}} \int \frac{(B_{\rm x}^{2} + B_{\rm y}^{2})}{2}\, dV.
    \label{eq:mag_xy}
\end{equation}

Here $\rm V_{\rm j}$ is the volume of the plasma column with dV = dX\,dY\,dZ. 

The growth rate ($\rm \eta_{\rm gr}$) of the kink instability is related to the development of the perturbation. It can be calculated from the evolution of the volume-averaged transverse (X-Y plane) kinetic energy in the linear phase as follows:
\begin{equation}
\eta_{\rm gr} = \frac{1}{E_{\rm peak}} \, \frac{dE_{\rm kin, xy}}{dt},
\end{equation}
where $E_{\rm peak}$ is the peak value of the transverse kinetic energy. 
The magnetic dissipation rate is associated with the non-linear development of instability and it can be calculated from the decaying of the volume-averaged transverse (X-Y plane) magnetic energy as follows:
\begin{equation}
\eta_{\rm diss} = \frac{1}{E_{0}} \, \frac{dE_{\rm mag, xy}}{dt},
\end{equation}
where $E_{0}$ is the initial transverse magnetic energy. 

\subsection{Numerical setup}\label{numerical_setup}

We perform the Relativistic MHD (RMHD) simulations of CD kink instability using the {\sc{PLUTO}} code \citep{Mignone_2007_PLUTO}. The numerical simulations are carried out by solving the following set of RMHD equations using Cartesian co-ordinates in their conservative form with rest mass density $\rho$, bulk velocity $\bvec{v}$, magnetic field $\bvec{B}$, and gas pressure $p_{\rm gas}$ \citep{Mignone_2013}.

\begin{equation}
    \frac{\partial (\gamma \rho)}{\partial t} + \nabla \cdot (\gamma \rho\, \bvec{v}) = 0,
\end{equation}

\begin{equation}
    \frac{\partial \bvec{m}}{\partial t} + \nabla \cdot \left[(w\gamma^{2}\bvec{v}\bvec{v} - \frac{\bvec{B}\bvec{B}}{4\pi} - \frac{\bvec{E}\bvec{E}}{4\pi}\right] + \nabla p_{\rm t} = 0,
\end{equation}

\begin{equation}
    \frac{\partial \bvec{\varepsilon}}{\partial t} + \nabla \cdot (\bvec{m} - \rho \gamma \bvec{v}) = 0,
\end{equation}

\begin{equation}
    \frac{\partial \bvec{B}}{\partial t} - \nabla \times (\bvec{v} \times \bvec{B}) = 0,
\end{equation}
where $\gamma$ is the Lorentz factor, the total momentum density $\bvec{m} = w\gamma^{2}\bvec{v} + \bvec{E} \times \bvec{B}/(4\pi)$, the electric field $\bvec{E}  = - \bvec{v} \times \bvec{B}$, $w$  is the enthalpy, the total pressure $p_{\rm t} = p_{\rm gas} + \frac{\bvec{B}^{2} + \bvec{E}^{2}}{8\pi}$ and the the total energy including thermal and magnetic energy $\varepsilon = w\, \gamma^{2} -\, p_{\rm gas} + \, \frac{\bvec{B}^{2} +\, \bvec{E}^{2}}{8\pi} - \rho \gamma $. An additional equation of a passive scalar quantity or a tracer ($\tau (x,\,y,\,z,\,t)$) is also included to differentiate between the \sa{column} material (where $\tau$ = 1) and the external medium (where $\tau$ = 0):
\begin{equation}
\frac{\partial \tau}{\partial t} + \bvec{v} \cdot \nabla \tau = 0.
\end{equation}

In the present work, we simulate a particular section of an AGN jet represented by a cylindrical plasma column. The numerical domain defined in Cartesian geometry has dimensions 4 $\times$ 4 $\times$ 6 with resolution $240 \times 240 \times 360$. 
We define the plasma column with radius $R_{\rm j} = 0.5$ and within this column we prescribe a constant velocity with Lorentz factor $\Gamma$ along $\hat{z}$. The azimuthal velocity inside the column is set to 0 indicating an initially non-rotating jet. The ambient medium outside of this column is static. The high resolution covering $R_{\rm j}$ with 30 grid points is essential to capture the transition of dynamical quantities at the column boundary.

\begin{figure*}
    \centering
    \includegraphics[scale=0.2]{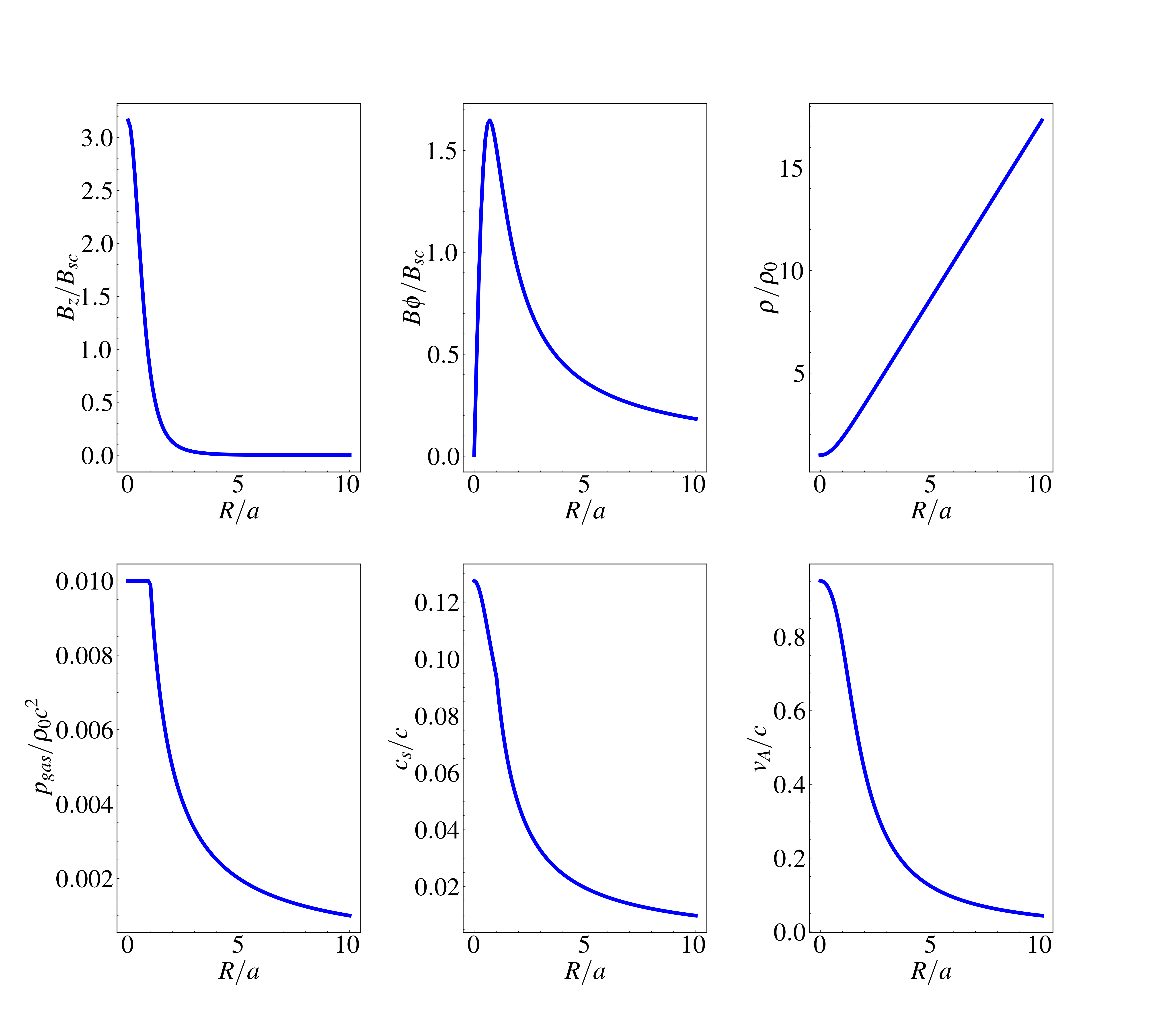}
    \caption{Radial profile of magnetic fields ($B_{z}$ $\&$ $B_{\phi}$), density ($\rho$), gas pressure ($p_{\rm gas}$), sound speed ($c_{\rm S}$) $\&$ Alfv\'en speed ($v_{\rm A}$) for the Ref case.}
    \label{fig:func_form}
\end{figure*}

The initial magnetic field is set using the force free condition defined below \citep{Anjiri_2014} :
\begin{equation}
    B_{\rm z}\frac{dB_{\rm z}}{dR} + \frac{B_{\rm \phi}}{R}\frac{d}{dR}(RB_{\rm \phi}) = 0,
    \label{eq:force_free}
\end{equation}
where, the magnetic field in the radial direction $B_{\rm r}=0$. 
The solution of the above equation can be expressed as \citep{Mizuno_2011}:
\begin{equation}
    B_{\rm z} = \frac{B_{0}}{\left[1 + \left(\frac{R}{a}\right)^{2}\right]^{\alpha}},
    \label{eq:Bz}
\end{equation}
\begin{equation}
    B_{\rm \phi} = \frac{B_{0}}{\left(\frac{R}{a}\right)\left[1 + \left(\frac{R}{a}\right)^{2}\right]^{\alpha}} \sqrt{\frac{\left[1 + \left(\frac{R}{a}\right)^{2}\right]^{2\alpha}-1-2\alpha \left(\frac{R}{a}\right)^{2}}{2\alpha-1}}.
    \label{eq:Bphi}
\end{equation}
In the above equations, R = $\sqrt{x^2 + y^2}$ is the radial position in a cylindrical coordinate system, $a = R_{\rm j}/2 = 0.25$ is the characteristic radius within the plasma coulmn at which the toroidal component of the magnetic field is maximum, $B_{0}$ is the magnetic field value at the axis, which is controlled through the magnetization parameter ($\sigma$), and $\alpha$ is the parameter that describes the radial profile of the pitch parameter defined as $P = r\frac{B_{\rm z}}{B_{\rm \phi}}$. \EC{The magnetic field distribution is put in the all simulation region and the radial profiles of the magnetic field components are provided in figure \ref{fig:func_form}.}
The magnetization parameter in the relativistic form is given by :
\begin{equation}
    \sigma = \frac{b^{2}}{\rho} = \frac{1}{\rho} \left[\frac{B^{2}}{\gamma^{2}} + (\bvec{v}.\bvec{B})^{2} \right],
\end{equation}
and at the initial time on the axis (R = 0), it reduces to :
\begin{equation}
    \sigma_{0} = \frac{B_{0}^{2}}{\rho_{\rm c}},
\end{equation}
where $b$ and $B$ correspond to the magnitude of total magnetic field in the co-moving and observer frame, respectively and $\rho_{\rm c}$ = 1.0 as the density on the axis in the non-dimensional units. \\

The non-uniform density profile is taken as $\rho = \frac{\rho_{\rm c}}{B}$. Initially, we consider a low gas pressure $p_{\rm gas}$ to maintain a low plasma beta and hence a high magnetization. The pressure profile is given as:
\begin{equation}
\label{eq:prs_distrb}
    p_{\rm gas} = 
\begin{cases}
    0.01,& \text{for } R < a \\
    0.01 \times \left(\frac{a}{R}\right),              & \text{elsewhere}
\end{cases}
\end{equation}

An ideal gas equation of state has been considered with $p_{\rm gas} = (\Gamma_{\rm ad} - 1)\rho e$, where $e$ is the specific internal energy density and the adiabatic index $\Gamma_{\rm ad}$ is set to 5/3.
In terms of light speed, the sound speed is $c_{\rm S}/c = \sqrt{\Gamma_{\rm ad} \, p_{\rm gas}/\rho h}$ and the Alfv\'en speed is given by $v_{\rm A}/c = \sqrt{B^{2}/(\rho h + B^{2})}$ where the specific enthalpy is given by $h = 1 + e/c^{2} + p_{\rm gas}/\rho c^{2}$.

To onset the kink instability, we perturbed the plasma column with a radial velocity in all regions given by \citep{Mizuno_2011}:
\begin{equation} \label{eq:pert}
    v_{\rm R} = \delta v \,\exp\left(-\frac{R}{R_{\rm j}}\right)\,\cos(m\theta)\,\sin\left(\frac{2 \pi zn}{L_{\rm z}}\right).
\end{equation}
We choose $m=1$ and different $n$ values (see table:\ref{tab:run_details}) as a specific choice for the kink mode instability and have adopted perturbation amplitude $\delta v = 0.01$, and $L_{\rm z} = 6$ is the length of the simulation box in the Z direction. 
In the output of the simulation, all the vector quantities are in the observer frame, whereas the scalar quantities are in the jet co-moving frame.

The simulation box is periodic in the Z direction, and the outflow boundary condition is imposed in the transverse X-Y direction. The periodic boundary condition allows us to study its temporal evolution, in particular the growth of the instability that is seeded with the above perturbation.
We have used the hlld Reimann solver, and constrained transport formalism \citep{Londrillo_2004_divB, Gardiner_2005_divB} to maintain divergence condition of the magnetic fields.

The simulations are done using dimensionless quantities and can be related to physical values using appropriate scales. 
We define three physical scales: the length scale L$_{\rm sc}$ = 0.1\,pc, the velocity of light c = $2.998\times 10^{10}$cm/s and the density at the initial time $\rho_{\rm 0}$ = $1.673 \times 10^{-24}$ gm/cm$^{3}$. 
The derived scaled unit for time is t$_{\rm sc} =  0.32$ years, for the pressure is $1.5 \times 10^{-3}\,\rm{dyne}\,\rm{cm}^{2}$ and for the magnetic field is B$_{\rm sc}$ =  1.374 $\times 10^{-1}$ Gauss respectively. 

\subsection{Parameter details}

Blazar jets are magnetized and highly relativistic with typical bulk Lorentz factor value $\Gamma$ = 5 or more \citep{Jorstad_2005}. The inner portion of the jet is generally lighter compared to the outer portion of the jet \citep{Walg_2013}. Even though both toroidal and poloidal magnetic fields are responsible for the formation and stability of the jet, it is the toroidal magnetic field component that enhances the kink instability effect \citep{Bromberg_2019}.

Our simulations are carried out with moderate Lorentz factors $\Gamma$ = 5 and 10 (Plasma column velocity V$_{\rm j}/c \approx$ 0.97 $\&$ 0.99 respectively) and have an increasing density profile ($\rho \approx R$).
It is expected that within the column radius (R$_{\rm j}$), the kink would move with the plasma and there will be a strong interaction between the growing kink and the flow \citep{Mizuno_2011}. Further, the pitch parameter is chosen to have a decreasing radial profile by considering $\alpha$ = 2.0. Having a decreasing pitch profile is more appropriate for generating kink in relativistic flow, suggesting an increasing toroidal magnetic field ($B_{\rm \phi}$) with the radius of the plasma column. 

\cite{Asada_Nakamura_2012} suggest that the jet maintains a parabolic structure up to $10^{5}$ times the Schwarzschild radius for a black hole of mass $\approx$ $10^{9}$ M$_{\odot}$. This implies the plasma column with diameter 0.1 pc would be situated at nearly $\sim 10$ parsecs away from the black hole. Thus, the plasma column considered in the present work would represent a portion of an AGN jet which is about few 10s of pc away from the central engine and will be magnetically dominated. We therefore chose two different co-moving magnetization values $\sigma_{0} = 1.0,$ and $10.0$ to analyze the effect of the magnetic field strength on the formation and evolution of the kink instability. The details of the different simulation setups are given in table \ref{tab:run_details}.

It is important to note that the choice of a radially declining gas pressure profile (Eq.~\ref{eq:prs_distrb}) refers to a configuration that is strictly not in a static state equilibrium. We have also performed our reference simulation with constant pressure $p_{\rm gas} = 0.01$ which ensures radial equilibrium and the corresponding result is discussed in appendix \ref{app:effect_bc}. In addition, we also study the effect of different transverse boundary conditions on the growth of the instability. We observe that the growth of the kink instability is not affected significantly irrespective of the choice of the initial pressure distribution and the transverse boundary condition. 
A detailed comparison of these cases is presented in appendix \ref{app:effect_bc}. \EC{Further, in addition to the pressure gradient force, the effect of the radial electric field is also not accounted for in equation \ref{eq:force_free} to obtain the equilibrium magnetic field (see equations \ref{eq:Bz} and \ref{eq:Bphi}). However, it does not affect the qualitative nature of the growth of the instability. A brief discussion on this is given in appendix \ref{sec:equi_app}.}

\begin{center}
\begin{table}
    \centering
    \begin{tabular}{|c|c|c|c|c|c|c|}
    \hline\hline
    Runs ID  & $\sigma_{0}$ & $\Gamma$  & $n$ & $\beta_{0}$ & v$_{A0}/c$   & t$_{\rm stop}/\rm t_{\rm sc}$ \\  [1.5ex]
    \hline\hline
    Ref\_n2 & 10.0 & 5.0 & 2 & 0.002 & 0.952 & 100 \\ [1.5ex]
    Ref\_n3 & 10.0 & 5.0 & 3 & 0.002 & 0.952 & 100 \\ [1.5ex]
    Ref\footnotemark[1] & 10.0 & 5.0 & 4 & 0.002 & 0.952 & 100 \\ [1.5ex]
    Ref\_n8 & 10.0 & 5.0 & 8 & 0.002 & 0.952 & 100 \\ [1.5ex]
    Ref\_s1 & 1.0 & 5.0 & 4 & 0.02 & 0.702 & 100 \\ [1.5ex]
    Ref\_g10 & 10.0 & 10.0 & 4 & 0.002 & 0.952 & 200 \\ [1.5ex]
    Ref\_g10\_s1 & 1.0 & 10.0 & 4 & 0.02 & 0.702 & 200 \\ [1.0ex]
    \hline\hline
    \end{tabular}
    \caption{All simulation runs details are given column wise as runs ID, magnetization value at the axis ($\sigma_{0}$), Bulk Lorentz factor ($\Gamma$), number of axial wavelengths ($n$) to be fitted inside the simulation box, the plasma beta parameter on the axis of jet ($\beta_{0} = \rm 2P_{0}/B^{2}_{0}$), the Alfv\'en speed ($v_{\rm A}$) and the time stamp at which the simulation stops (t$_{\rm stop}$) respectively. In all the cases the pitch profile parameter ($\alpha$) is 2.0 and c$_{s0}/c$ is 0.127. These values are given at the initial time of the simulation.}
    \label{tab:run_details}
\end{table}  
\footnotetext[1]{Two simulations named as Ref\_A1 and Ref\_A2 have been performed with different initial and boundary condition. The details of the simulations and the corresponding results are provided in the appendix \ref{app:effect_bc}. \EC{Additionally, two more simulations named as Ref\_B1 and Ref\_B2 have also been performed to investigate the equilibrium configurations. See section \ref{sec:equi_app} for a more detailed discussion.}}
\end{center}

\subsection{Non-thermal emission modeling}\label{emission_setup}

To observe the effect of the kink instability on the emission process of jets, we developed a post-processing code to estimate the synchrotron emission using the ray-tracing method. All relativistic effects are taken into consideration to account for the relativistic beaming and boosting along with light travel effect. The inputs are the fluid variables obtained from the simulations done by using the {\sc{PLUTO}} code along with the viewing angle and the parameters that formulate the particle spectrum.
This code calculates the synchrotron emissivity in the observer frame as an output. The synchrotron emission is primarily due to the non-thermal particles with a single power-law particle distribution with index $p$. 
In our approach, each grid cell acts as a single emitting blob. However, the only limitation is that the particle distribution remains unchanged with time. 

We have considered the total energy density to be a fraction of thermal energy density, i.e., 
\begin{equation}
    \int \gamma^{\prime}_{\rm e} \, N^{\prime}(\gamma^{\prime}_{\rm e})\, d\gamma^{\prime}_{\rm e} = \zeta \varepsilon_{\rm th},
\label{eq:thermN}
\end{equation}
where $\zeta$ = 0.05 and $\gamma^{\prime}_{\rm e}$ is the electron Lorentz factor.
The total synchrotron emissivity in the co-moving frame for a particular frequency ($\nu^{\prime}$) and direction ($\hat{n}^{\prime}$) is calculated by integrating the product of the power emitted by a single electron with the particle distribution,
\begin{equation}
    J_{\rm syn}^{\prime}(\nu^{\prime}, \hat{n}^{\prime}) = \int P^{\prime}(\nu^{\prime},\gamma^{\prime}_{\rm e})\, N^{\prime}(\gamma^{\prime}_{\rm e})\, d\gamma^{\prime}_{\rm e}
\end{equation}

For a power-law distribution, it can be expressed as \citep{POLARIS_2019}:
\begin{equation}
\begin{split}
    J_{\rm syn}^{\prime}(\nu^{\prime}, \hat{n}^{\prime}) = \frac{\nu^{\prime}_{\rm G} \sin \theta (p-1) 3^{\rm p/2}}{\gamma^{\prime\,1-p}_{\rm min} - \gamma^{\prime\,1-p}_{\rm max}}
   \left( \frac{\nu^{\prime}}{\nu^{\prime}_{\rm G} \sin\theta} \right)^{\frac{\rm 1 - p}{2}} \frac{n_{\rm e}^{\rm NT} e^{2}}{c 2^{\frac{\rm p+3}{2}}}  \\ \times \int_{x_{\rm 1}}^{x_{\rm 2}} F(x) x^{\frac{\rm p-3}{2}} dx,
\label{eq:SyncEms}
\end{split}
\end{equation}
where $\nu^{\prime}_{\rm G} = \frac{eB^{\prime}}{2\pi m_{\rm e}c}$ is the gyro-frequency of an electron \citep{Longair2011} in the co-moving frame, $\gamma^{\prime}_{\rm min}$ and $\gamma^{\prime}_{\rm max}$ are the limits of the electron energies, $n_{\rm e}^{\rm NT}$ is the non-thermal particle number density, $\theta$ is the angle between $\bvec{B^{\prime}}$ and $\bvec{\hat{n}^{\prime}}$, and $\rm F(x)$ can be found by integrating the modified Bessel function of the order of $5/3$ in the following manner:
\begin{equation}
    F(x) \equiv x \int_{x}^{\infty} K_{5/3}(\xi)d\xi,
\end{equation}
where $x \equiv \frac{\nu^{\prime}}{\nu^{\prime}_{c}}$ and $\nu^{\prime}_{\rm c}$ is the critical frequency of synchrotron emission for a single electron given by
\begin{equation}
    \nu^{\prime}_{\rm c} = \frac{3}{2} \gamma^{\prime\,2} \nu^{\prime}_{\rm G} \sin\theta.
\end{equation} 

The emissivity in the co-moving frame can be transformed into the observer's frame as follows
\begin{equation}
    J_{\rm syn}(\nu,\, \hat{n}) = \delta^{2} J_{\rm syn}^{\prime}(\nu^{\prime},\, \hat{n}^{\prime}).
\end{equation}
The quantities in the comoving frame such as $\nu^{\prime}, \hat{n}^{\prime}, B^{\prime}$ also have to be transformed and can be expressed as functions of $\nu, \hat{n},$ and $B$ \citep[see e.g.,][]{DelZanna_2006, Vaidya_2018}.
The Doppler factor $\delta$ is obtained from
\begin{equation}
    \delta(\beta, \hat{n}) = \frac{1}{\Gamma(1 - \beta . \hat{n})},
    \label{eq:dop_Fact}
\end{equation}
with $\Gamma$ as the bulk Lorentz factor and $\beta$ as the bulk velocity in terms of $c$.

We use the radiative transfer equation to calculate the intensity.
\begin{equation}
    \frac{dI_{\rm \nu}}{ds} = -\alpha_{\rm \nu}I_{\rm \nu} + J_{\rm \nu},
\end{equation}
where $\alpha_{\rm \nu}$ and $J_{\rm \nu}$ are the absorption and emission coefficients. For simplicity, we consider an optically thin regime implying
\begin{equation}
    I_{\rm \nu} = \int J_{\rm \nu} ds,
\end{equation}
where $ds$ is the infinitesimal distance along line of sight (los) traveled by the emitting photon. The synchrotron flux density is then calculated using the following formula:
\begin{equation}
    F_{\rm \nu} = \int I_{\rm \nu} d\Omega,
\end{equation}
where the subtended solid angle is given by $d\Omega$ = $\frac{dA}{D^{2}}$, where dA (dx $\times$ dy) is the area of one grid cell, and $D$ is the distance between the source and the observer. 
In this work, we have adopted a reference distance $D = 7.9$ Mpc between the source and the  observer. By considering the length scale and the grid resolution, we obtain the solid angle as
\begin{equation}
    \frac{d\Omega}{4\pi} = 3.51 \times 10^{-21} \left(\frac{D}{7.9 \rm Mpc}\right)^{-2}.
\end{equation}

The simulated flux density is scaled with ${F_{\nu_{\rm sc}}}$, defined as \citep{Borse_2020}
\begin{equation}
    {F_{\rm \nu_{\rm sc}}} = \frac{E_{\rm sc}c}{{r_{\rm L}}^3\nu_{\rm sc}} = 4\pi I_{\rm \nu_{\rm sc}} = 4\pi j_{\rm \nu_{\rm sc}} r_{\rm L},
\end{equation} 
where, $E_{\rm sc} = \gamma_{\rm sc}m_{e}c^{2}$ is the energy scale with $\gamma_{\rm sc} = 1$, and frequency is scaled in the units of Larmor frequency $\nu_{sc} = \nu_{\rm G} \approx 8.84 \times 10^{5}$ Hz. Further, $r_L$ is the Larmor radius for highly relativistic electrons, with Lorentz factor $\gamma_{\rm max}$, used in the calculation of synchrotron emission. The values of $\nu_{sc}$ and $r_L$ are estimated using the initial magnetic field strength defined at the axis of the plasma column (i.e., $B_z(t=0)$ = 0.31 Gauss). 
For the chosen set of physical scales, the value of the flux scale in physical units is ${F_{\nu_{\rm sc}}} \approx 4.765 \times \left(\frac{\gamma_{\rm max}}{10^{6}}\right)^{-3}$ ergs s$^{-1}$ cm$^{-2}$ Hz$^{-1}$. 

The idealized value of the flux is obtained from the above post-processing module on RMHD simulations. The temporal evolution of such an idealized flux value for a particular frequency $\nu$, and the defined line of sight is referred to as a synthetic light curve.
With an aim to incorporate standard noise during actual observations, we add random error to the values obtained for the synthetic light curve. 
The error is added by setting a target signal to noise ratio (SNR). Using this user-defined value of SNR = $\frac{P_{\rm signal}}{P_{\rm noise}}$, the average power of noise ($P_{\rm noise}$) is calculated. A random normal error is generated by taking the mean as zero and the square root of the average power of noise as standard deviation and added to the idealized flux value at each time to generate a more realistic light curve.

\section{Results of dynamics}\label{results}

We describe the results obtained from the dynamical study of the magnetized jets for the Ref case in section \ref{sec:Ref_case_dynamics}. Further, the results obtained from studying different parameters such as $n$, $\sigma$ and $\Gamma$ is explained in section \ref{sec:parameter_study}.

\subsection{Results from the reference case}\label{sec:Ref_case_dynamics}

\begin{figure*}
     \centering
     \includegraphics[scale=0.41]{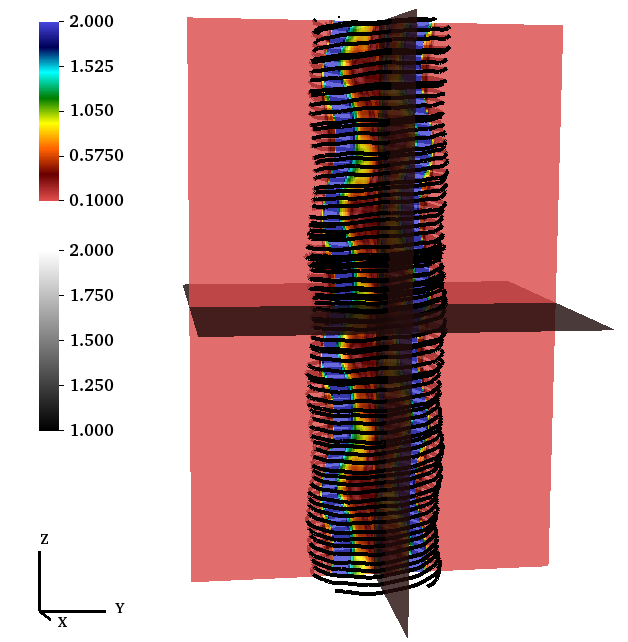}
     \includegraphics[scale=0.4]{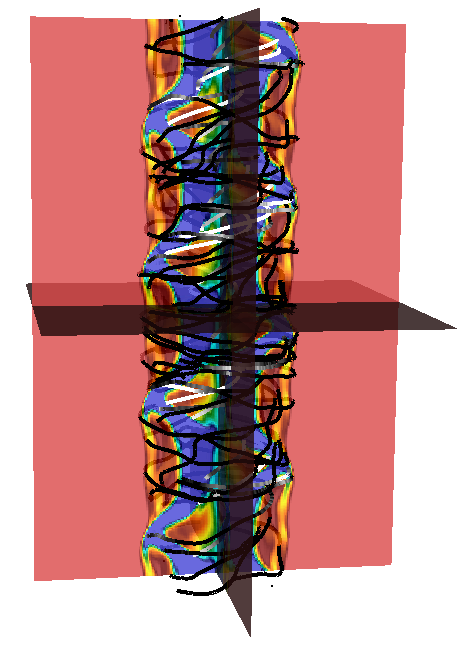}
     \includegraphics[scale=0.405]{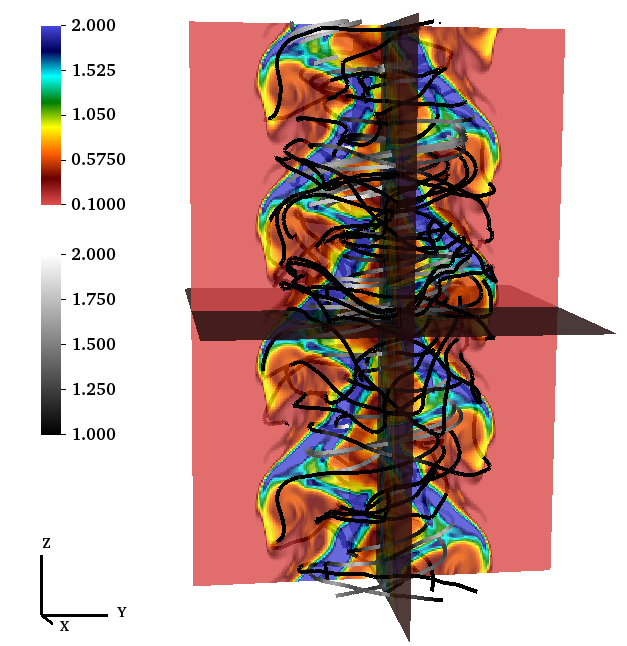}
     \includegraphics[scale=0.4]{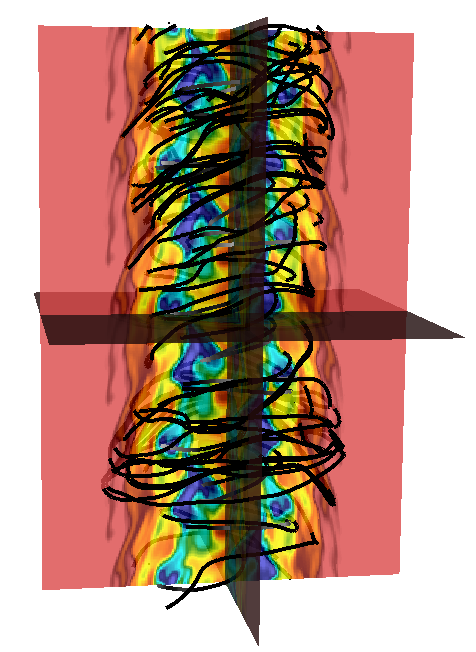}
     \caption{Time evolution of a three slice representation of the plasma column density at $t/t_{\rm sc} = 30, 50, 70,$ and $90$ after kink mode perturbation for the Ref case. The colorbars show the plasma density and magnetic field lines strength. The cuts are taken at the planes (X = 0, Y = 0, Z = 0.5$L_{\rm z}$)}
    \label{fig:Ref_dynamics} 
\end{figure*}

Figure \ref{fig:Ref_dynamics} shows the time evolution of the plasma column density and the magnetic field lines for the Ref case in the computational units. As time evolves, we observe a growth of the $m$ = 1 mode perturbation resulting in distortion of the plasma column. The density gets concentrated near the kinked portion and magnetic field lines also get tangled with the evolution of the instability. In the non-linear phase of its growth, the plasma column continues to propagate in the transverse direction making the structure more twisted and helical. The jet density in the back-ground is over-plotted with the magnetic field lines, where the gray colorbar represents the magnetic field strength.
As the plasma column evolves, we see a mixing of ambient material at the column boundary due to the velocity shear. Since the helical magnetic field is a dominant component in the Ref case, it suppresses the formation of vorticity at the boundaries and the jet remains stable \citep{Malagoli_1996, Appl_1992, Baty_2002, Borse_2020}.

\begin{figure*}
    \centering
    \includegraphics[scale=0.45]{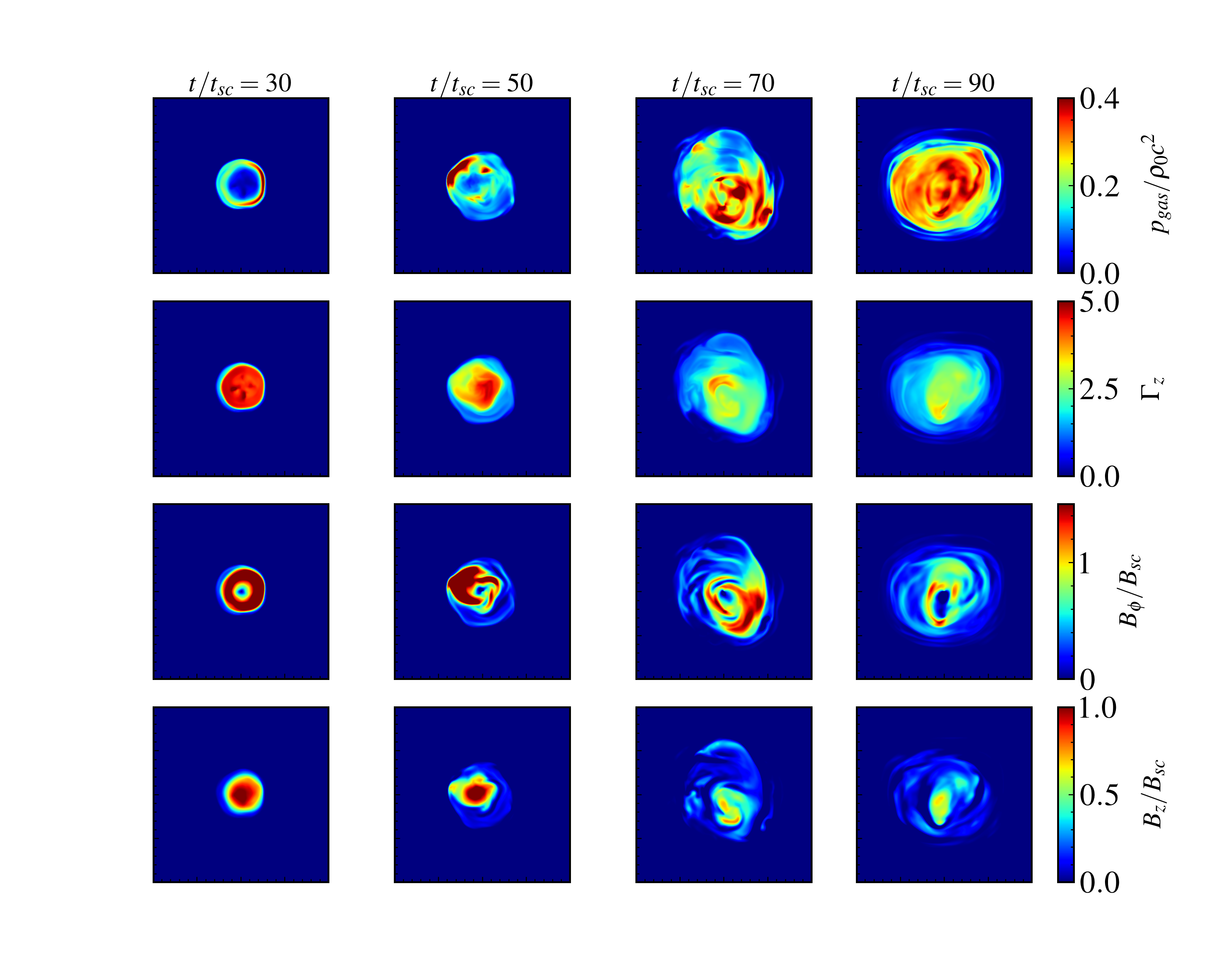}
    \caption{Time evolution of the plasma column pressure, the axial Lorentz factor, the azimuthal and poloidal magnetic field components (row-wise) in the X-Y plane of the Ref case at $t/t_{\rm sc} = 30, 50, 70,$ and $90$.}
    \label{fig:Ref_4Slices}
\end{figure*}

In figure \ref{fig:Ref_4Slices}, we have shown the 2D slices of the plasma column at the Z = 0.5$L_{\rm z}$ plane representing the gas pressure ($p_{\rm gas}/\rho_{0}c^{2}$), the axial Lorentz factor ($\Gamma_{\rm z}$), azimuthal ($B_{\rm \phi}/B_{\rm sc}$), and the axial magnetic field ($B_{\rm z}/B_{\rm sc}$) components at different evolutionary stages. These distributions give a view of the plasma column parallel to the axis of it for the Ref case. The pressure distribution indicates that the plasma column is expanding with the evolution of the instability. At the early time $t/t_{\rm sc} = 30$, the pressure is largely distributed at the shear boundary. However, with the expansion of the plasma column and the development of the instability, it gets distorted. The second and third row represent the distribution of axial velocity (i.e., the axial Lorentz factor $\Gamma_{\rm z}$), and the azimuthal component of magnetic field, that clearly indicates the movement of the kinked portion of the plasma column. The twisting is also clearly visible as the perturbation starts affecting the column from the shear interface and $B_{\rm \phi}$ is also strong enough near the shear boundaries. The fourth row of figure \ref{fig:Ref_4Slices} reflects the strength of the axial component of the magnetic field, that maintains the stability of the column. At 
$t/t_{\rm sc} = 30$, the value of $B_{\rm z}/B_{\rm sc}$ is maximum and highly concentrated near the axis of the column. At later times, as the instability evolves, the effect of the stabilizing is reduced due to dissipation of the axial component.

The morphological variations observed due to the onset of the instability can be quantified by defining the average position of the barycenter $\bar{r}(z)$ as a function of height $z$ \citep{Mignone_2010, Mignone_2013}, such as
\begin{equation}
\bar{r}(z) = \sqrt{\bar{x}(z)^2 + \bar{y}(z)^2}, 
\label{eq:barycentre}
\end{equation}
where
\begin{equation}
    \bar{x}(z) = \frac{\int x\,Q(x, y, z)\,dx\,dy}{\int Q(x, y, z)\,dx\,dy},\, \bar{y}(z) = \frac{\int y\,Q(x, y, z)\,dx\,dy}{\int Q(x, y, z)\,dx\,dz},
\end{equation}
where $Q(x, y, z)$ is any flow quantity such as the density, Lorentz factor etc. used as the weight factor to include or exclude certain regions as per the requirement.
In our analysis, we consider density as the weighing factor as the focus is on the kinked portion and it is expected that the matter would get accumulated at the twisted or bent portion of the plasma column.

\begin{figure*}
    \centering
    \includegraphics[scale=0.16]{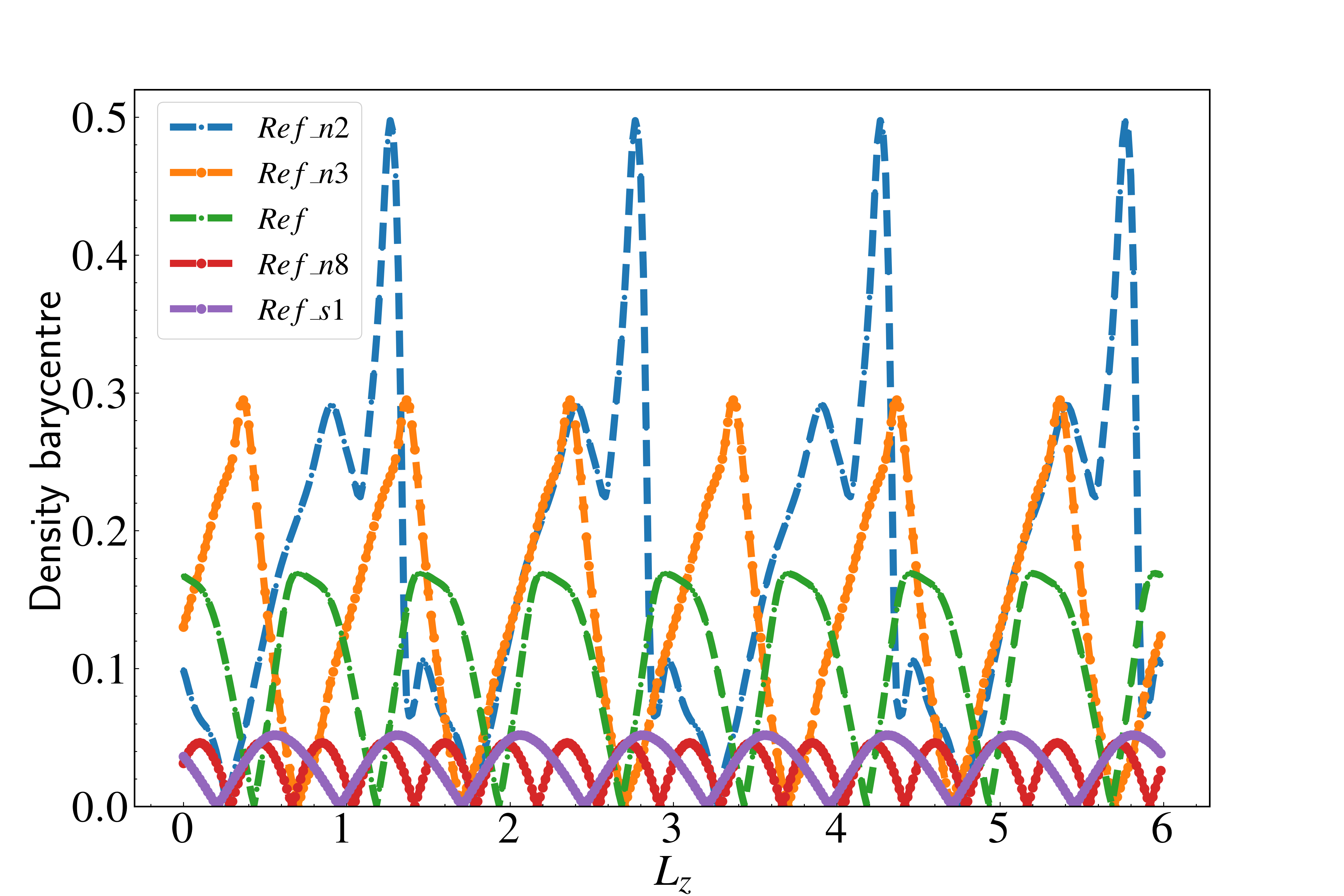}
    \includegraphics[scale=0.16]{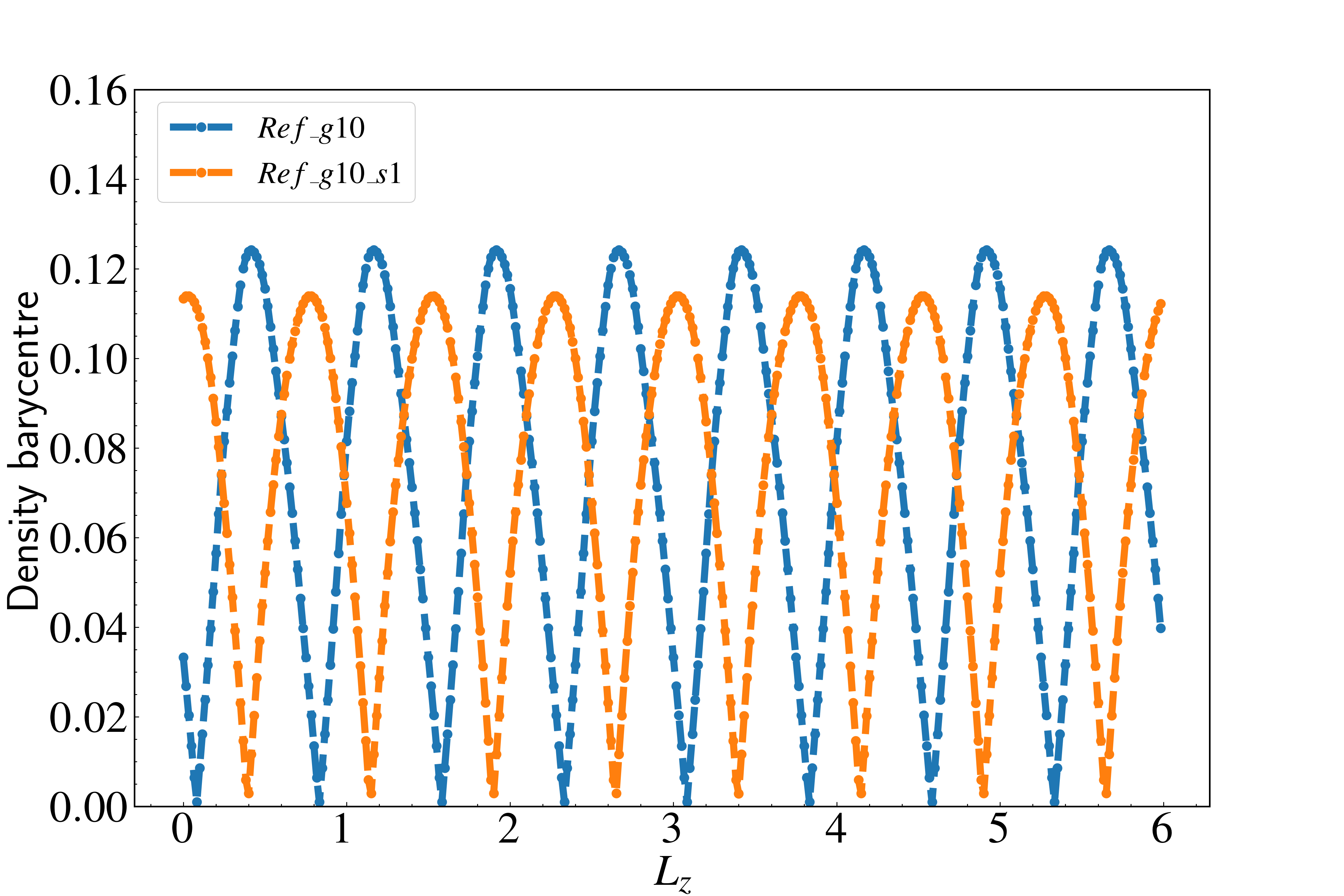}
    \caption{Density barycentre plots for the runs Ref\_n2, Ref\_n3, Ref, Ref\_n8, Ref\_s1 at $t/t_{\rm sc} = 50$ are in the left panel and for the runs Ref\_g10, Ref\_g10\_s1 at $t/t_{\rm sc} = 95$ are in the right panel.}
    \label{fig:rho_bary} 
\end{figure*}

Figure \ref{fig:rho_bary} shows the density barycentre motion of the jet for all the runs, demonstrating the impact of different $n$, $\sigma$, and $\Gamma$ on the evolution of the instability. Based on the form of the radial perturbation, the motion of the density barycentre shows similar pattern where the number of peaks is twice the number of axial wavenumbers that fit into the simulation box. The green dashed line in the left panel of figure \ref{fig:rho_bary} shows the distribution of the averaged density on each slice of the plasma column for the Ref case at $t/t_{\rm sc} = 50$. At this time stamp, the instability has evolved enough to form a helical structure. As a result, the variation in the position of the density barycentre is visibly periodic along the axis of the plasma column.

\begin{figure*}
    \centering
    \includegraphics[scale=0.4]{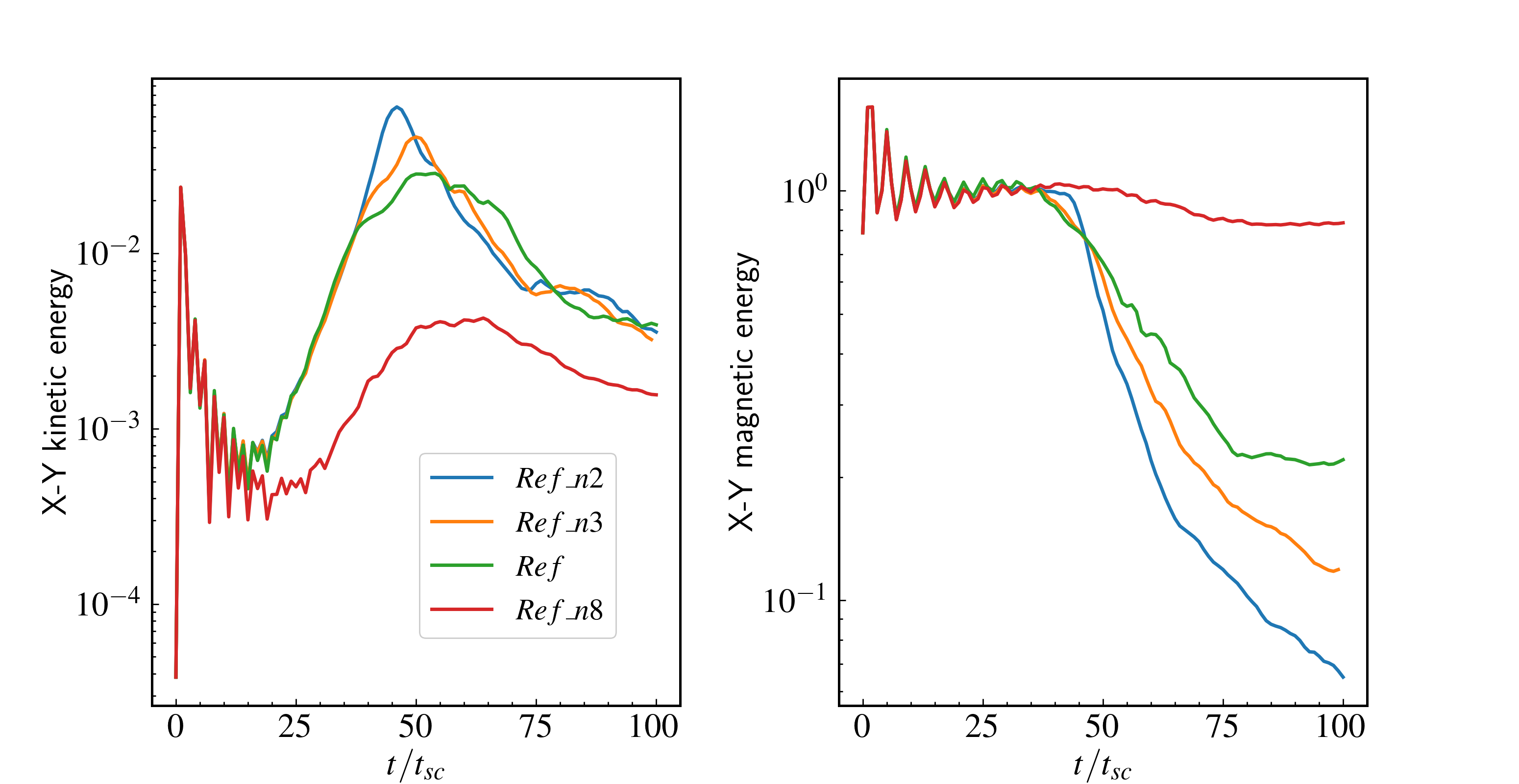}
    \caption{Time evolution of the volume averaged quantities for the cases Ref\_n2, Ref\_n3, Ref, and Ref\_n8.}
    \label{fig:energy_g5}
\end{figure*}

As an indicator of the growth of the instability, we have plotted the time evolution of averaged quantities defined in section \ref{sec:CD_mode} given in equations \ref{eq:kin_xy} and \ref{eq:mag_xy}.
Figure \ref{fig:energy_g5} shows the time evolution of $E_{\rm kin,\rm xy}$ and $E_{\rm mag,\rm xy}$ for the cases Ref\_n2, Ref\_n3, Ref, and Ref\_n8.
The temporal evolution of the transverse kinetic and magnetic energy is opposite to each other. For the Ref case, the transverse kinetic energy increases till $t/t_{\rm sc} = 50-60$ at the expense of the magnetic energy, representing the linear growth phase of the instability. 

\subsection{Results from parameter study}\label{sec:parameter_study}
In this section, we provide a comparative study of the different parameters that we considered in our simulations and their effect on the evolution and growth of the instability.

\begin{itemize}

    \item Effect of axial wavenumber ($\rm n$): In this work, we have considered four different wavenumbers to understand their effect on the dynamics of the plasma column. The displacement of the density barycentre is maximum for Ref\_n2 and the displacement decreases with increase of $n$, by keeping the other parameters the same. The density barycentre displacement curve exhibits multiple peaks for the Ref\_n2 case, indicating the fragmentation of the column material. The multiple high-density blobs in the X-Y slices of the column lead to multiple peaks as the displacement is calculated by considering density as a weight factor. As the instability evolves, the plasma column expands, and the density gets distributed in a chaotic manner. However, in the Ref\_n8 case, the instability development is comparatively less. This is evident from low amplitude quasi-periodic variation of the density barycentre displacement as a function of height Z (red dashed line in Fig.~\ref{fig:rho_bary}). 
    
    In the linear growth phase, the behavior of developing kink is almost the same for different $n$ values with different peak value of $E_{\rm kin,\rm xy}$. The transverse magnetic energy, $E_{\rm mag,\rm xy}$ gradually decreases in the early linear growth phase, then depending upon the growth of kink, it exhibits a rapid or slower decrease into the non-linear phase. The later behaviour is same for $E_{\rm kin,\rm xy}$ curve as well. The rapid decrease of $E_{\rm mag,\rm xy}$ for the Ref\_n2 case shows the highest kink growth rate compared to cases with higher $n$ values.

    \item Effect of Lorentz factor ($\Gamma$): On comparing runs with different values of $\Gamma$ (Ref and Reg\_g10), we find that the density barycentre displacement $\bar{r}(z)$ is less for the $\Gamma$ = 10 case with the same $n$ value. 
    This is consistent with the fact that kink instability takes longer time to develop for high speed jets.

    From the evolution of averaged energies, the linear growth regime for the Ref\_g10 case shows similar qualitative behaviour as the $\Gamma = 5$ case. However, for high speed jets, the linear growth phase extends up to $t/t_{sc} \approx 80$ (plot not shown). The peak value of the transverse kinetic energy is also less by an order of magnitude as compared to the cases with $\Gamma$ = 5. This is due to the moderate evolution of the instability within the computational timescale for the high speed case. 
    Also the dissipation of transverse magnetic energy $E_{\rm mag,\rm xy}$ is slower as compared to the $\Gamma$ = 5 run with the same $n$ value because of less and delayed growth of instability.

    \item Effect of magnetization ($\sigma$): \EC{In our work, we have performed two simulations with lower $\sigma$ configurations with $\Gamma$ = 5 and 10 and $n$ = 4 (Ref\_s1 and Ref\_g10\_s1). In both cases, we have a radially decreasing pitch profile that ensures a stronger toroidal component of the magnetic field  compared to the poloidal component. 
    Such a magnetic field configuration promotes the growth of kink instability based on the value of flow Alfv\'enic Mach number. We have calculated the density-weighted averaged Alfv\'enic Mach number ($M_{\rm A}$) and have shown its evolution in figure \ref{fig:machNumber}. For the Ref\_s1 and Ref\_g10\_s1 cases, $M_{\rm A}$ is found to be within a range of 0.6-1.4, and 0.4-1.4 respectively, within our simulation time scale, indicating the plasma column to be in the trans-Alfv\'enic regime. In these cases, mixing of both kink and KHI is expected as the $M_{\rm A}$ values vary from initially in the sub-Alfv\'enic to later in the  trans-Alfv\'enic regime compared to the Ref case where the column is sub-Alfv\'enic in nature (see figure \ref{fig:machNumber}) and subjected only to the kink instability. Due to the variable Alfv\'enic nature and the presence of a stronger toroidal magnetic field component compared to the poloidal component, the growth of KHI in the $\sigma$ = 1 cases is suppressed compared to the growth of the kink \citep{Baty_2002}. In addition, in comparison to the higher $\sigma$ cases, a stalled growth of the kink mode is apparent for the lower $\sigma$ cases since the resultant magnetic field strength is reduced. As a result, the plasma column is not distorted enough by the perturbation for the lower $\sigma$ cases. Further, the distribution of the quantities such as density, pressure, Lorentz factor, etc., is concentrated near the axis of the column, indicating a slow growth of the kink instability. Thus, the density barycentre displacement is also less for the Ref\_s1 and Ref\_g10\_s1 cases compared to the Ref case (see figure \ref{fig:rho_bary}).}
    
    \begin{figure}
        \centering
        \includegraphics[scale=0.26]{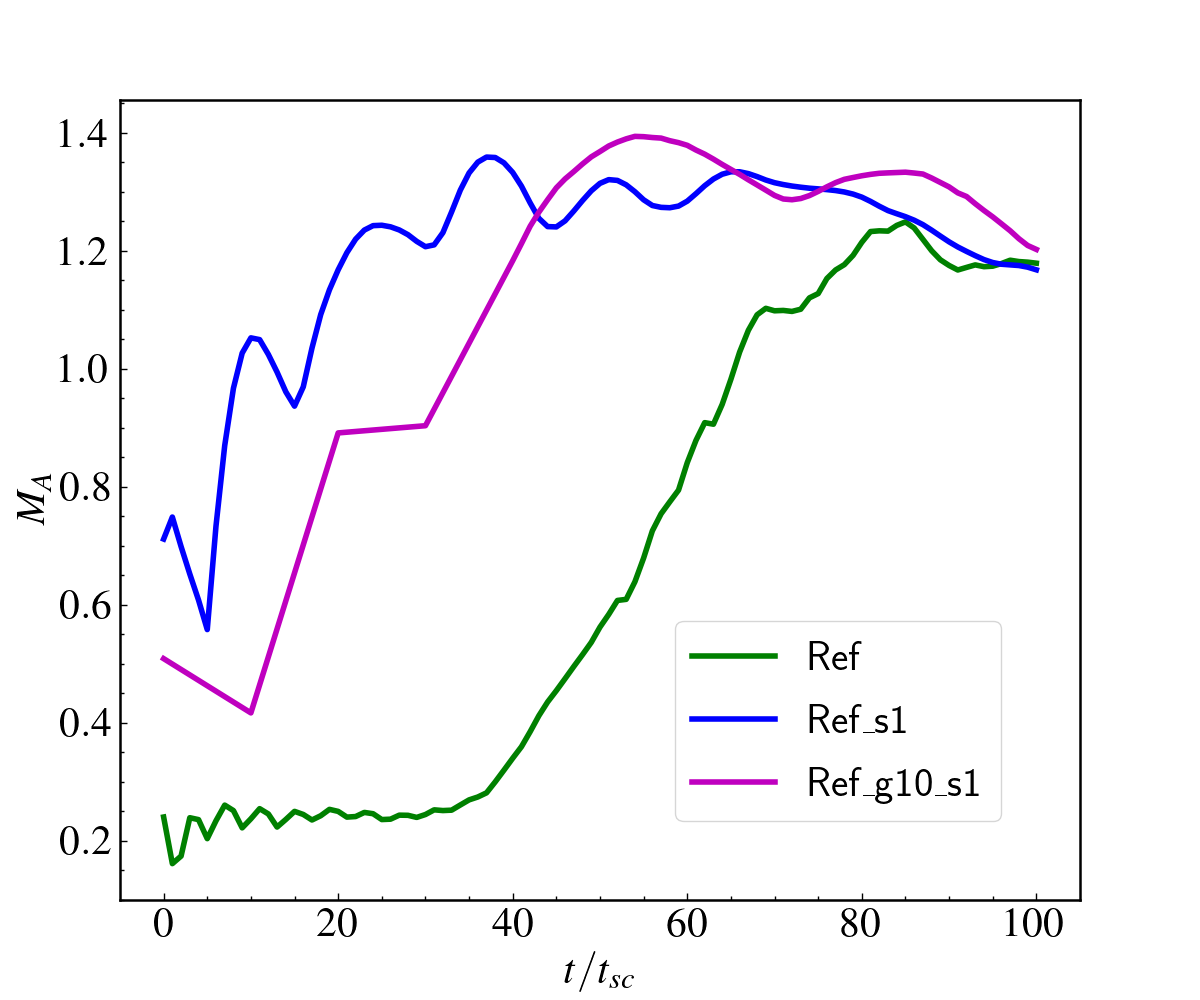}
        \caption{\EC{Time evolution of $M_{\rm A}$ for Ref, Ref\_s1 and Ref\_g10\_s1 cases.}}
        \label{fig:machNumber}
    \end{figure}
    
    \EC{In the lower magnetized cases, the temporal evolution of energies does not follow the same trend as that of highly magnetized cases since the nature of the instabilities is not similar compared to the $\sigma$ = 10 cases. In the $\sigma$ = 1 cases, (Ref\_s1 and Ref\_g10\_s1), the volume averaged transverse kinetic ($E_{\rm kin,\rm xy}$) and magnetic energy ($E_{\rm mag,\rm xy}$) decreases (plot not shown). The decay in the $E_{\rm kin,\rm z}$ curve is faster for the higher $\Gamma$ case due to a larger shear compared to the lower $\Gamma$ case. In the present work, we are mainly interested in highly magnetized jets that are subjected to the current driven kink mode instability and its effect on the dynamical and emission properties.}
    
\end{itemize}

\textbf{Growth rate ($\eta_{\rm gr}$) and dissipation rate ($\eta_{\rm diss}$):} 
As the high magnetization cases are more prone to experience the kink instability, the linear growth rate and the dissipation rate have
been calculated for the $\sigma$ = 10 cases by the formulations given in section \ref{sec:CD_mode} and provided in table~\ref{tab:Grate}. The growth rate for the Ref case is found to be $\approx$ 0.03, and for the Ref\_n2 case, $\approx$ 0.08, which is the maximum \EC{among the runs considered here}. It is significantly less in the case of Ref\_g10 compared to $\Gamma$ = 5 cases as the instability takes a longer time to set in due to the faster motion of the plasma column.
\EC{It should be noted that, in this work, we have shown the results for $n$ = 2, 3, 4 and 8 with all the other physical parameters being the same and among these cases, Ref\_n2 shows the highest kink growth rate. By using the formulations obtained from linear theory provided in section~\ref{sec:CD_mode} \citep{Appl_2000}, we obtained $k_{\rm max}$ $\approx$ 1.05 and $\eta_{{\rm gr}_{\rm max}}$ $\approx$ 0.18 for the considered pitch value $P_0 = 0.707$.
With the adopted domain size of axial length $L_{\rm z}$ = 6, the wave-number corresponding to $n=1$ would be 2$\pi n_{\rm max}/L_{\rm z}$ $\approx$ 1.04 and therefore will have the maximum growth following the linear analysis. To verify the same, we carried out an auxiliary run with $n=1$ (plot not shown) and quantified the growth rate of $0.1$ from the volume average kinetic energies. This estimate is higher than that obtained from $n = 2$ and also consistent with the value obtained from the linear analysis. 
The growth rate values with changing the value of $n$ obtained from the magnetic and kinetic energies, therefore, satisfy the trend derived from the linear theory analysis.} We also estimated the dissipation rate from the non-linear evolution of the transverse magnetic energy i.e. from the slope of the decay of magnetic energy. We found similar values of $\eta_{\rm gr}$ and $\eta_{\rm diss}$ and the trend for different runs remains the same. \EC{In addition, as the kink structure advects along with the jet with the evolution of the instability, we estimated the advection velocity of the kink and it was found to be $\approx$ (0.85-0.88)c for different $n$ values in the $\sigma$ = 10 cases that correspond to a Lorentz factor of $\approx$ (1.8-2.1).}

\begin{center}
\begin{table}
    \centering
    \begin{tabular}{|c|c|c|c|c|c|c|}
    \hline\hline
     Runs ID  & Ref\_n2  & Ref\_n3  & Ref  & Ref\_n8  & Ref\_g10   \\ [1ex]
     \hline
    $\eta_{\rm gr}$ & 0.08 & 0.044  & 0.033  & 0.006  & 0.008  \\ [1ex]
    \hline
    $\eta_{\rm diss}$ &  0.054 & 0.04 & 0.026 &  0.0061 & 0.004 \\ [1ex]
     \hline\hline
    \end{tabular}
    \caption{Kink linear growth rate ($\eta_{\rm gr}$) and magnetic energy dissipation rate ($\eta_{\rm diss}$) for the runs with $\sigma$ = 10.}
    \label{tab:Grate}
\end{table}
\end{center}

\section{Helical jet model}\label{helical_jet}

\subsection{Effect of relativistic boosting and viewing angle}\label{sec:Sync_plane}

The spectral and timing properties such as the long-term flux variation along with the optical outburst of CTA 102 are well explained by \cite{Raiteri_2017} with an in-homogeneous twisted jet model: the helical jet model, just with the variation of the viewing angle. According to this model, as the jet structure is dynamic, different regions of the helical jet (plasma column) have different orientations in time with respect to the line of sight of the observer. Therefore, the emission is more (less) enhanced when the region is better (worse) aligned with the specific line of sight.

In our case, the synchrotron emission is primarily due to the non-thermal particles with a single fixed power-law distribution with a power-law index $p$ = 3. This is a typical value chosen to capture magnetic reconnection that could be triggered by the kink instability and a viable energy dissipation mechanism and particle acceleration \citep{Bodo_2020}. The minimum and maximum energies of the electrons are taken to be $\gamma_{\rm min}$ = $10^{2}$ and $\gamma_{\rm max}$ = $10^{6}$. The emission maps are generated at an observing frequency of $\nu_{\rm obs}/\nu_{\rm sc}$ = $5.09 \times 10^{8}$, that corresponds to emission in optical (R-Band). The direction dependence along with the contribution of different parameters on the emission signatures are studied in the following section.

\begin{figure}
    \centering
    \includegraphics[scale=0.6]{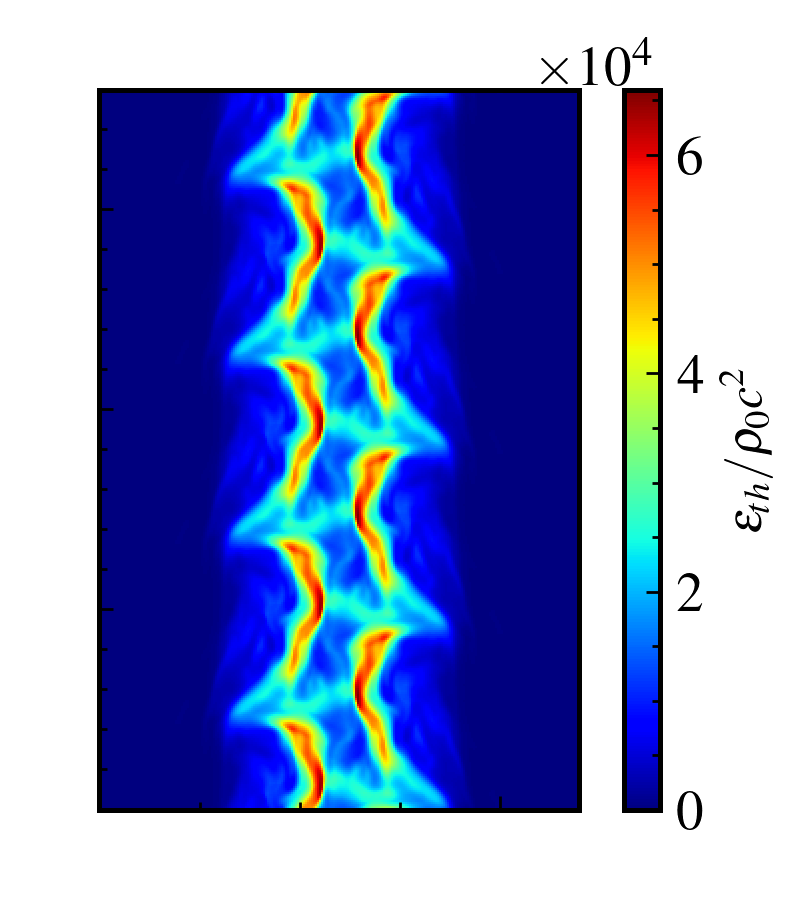}
    \caption{X-Z cut of thermal energy density normalized to its scaled value, for the Ref case at $t/t_{\rm sc}$ = 70.}
    \label{fig:ThEng_XZ}
\end{figure}

\begin{figure*}
    \centering
    \includegraphics[scale=0.5]{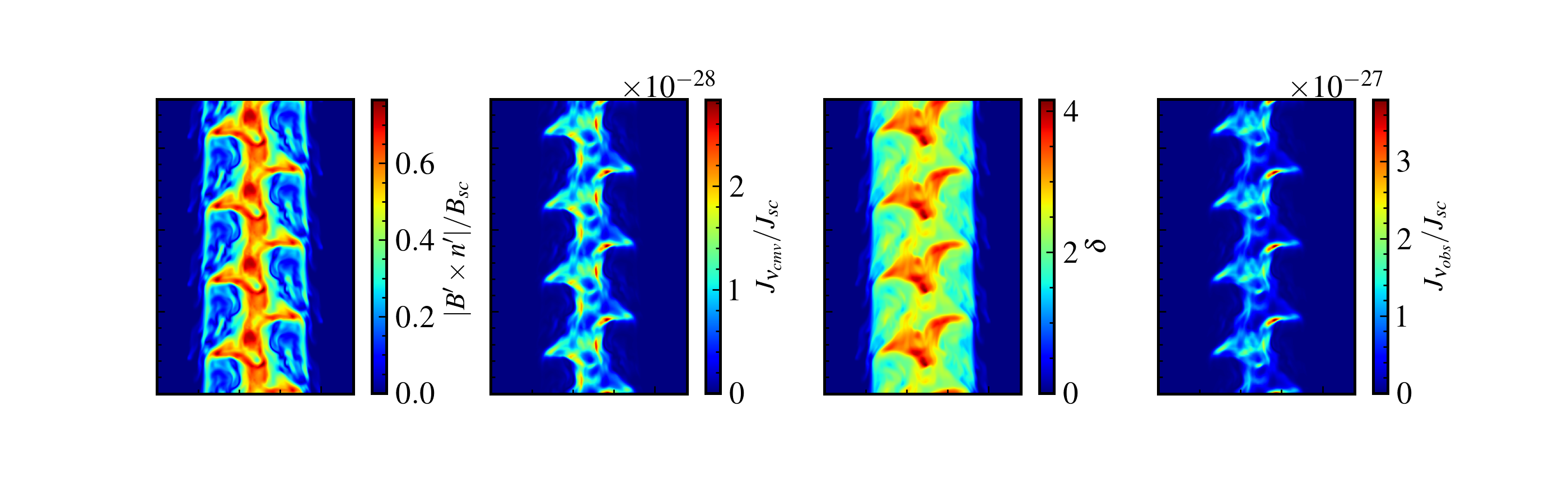}
    \vskip-2ex
    \includegraphics[scale=0.5]{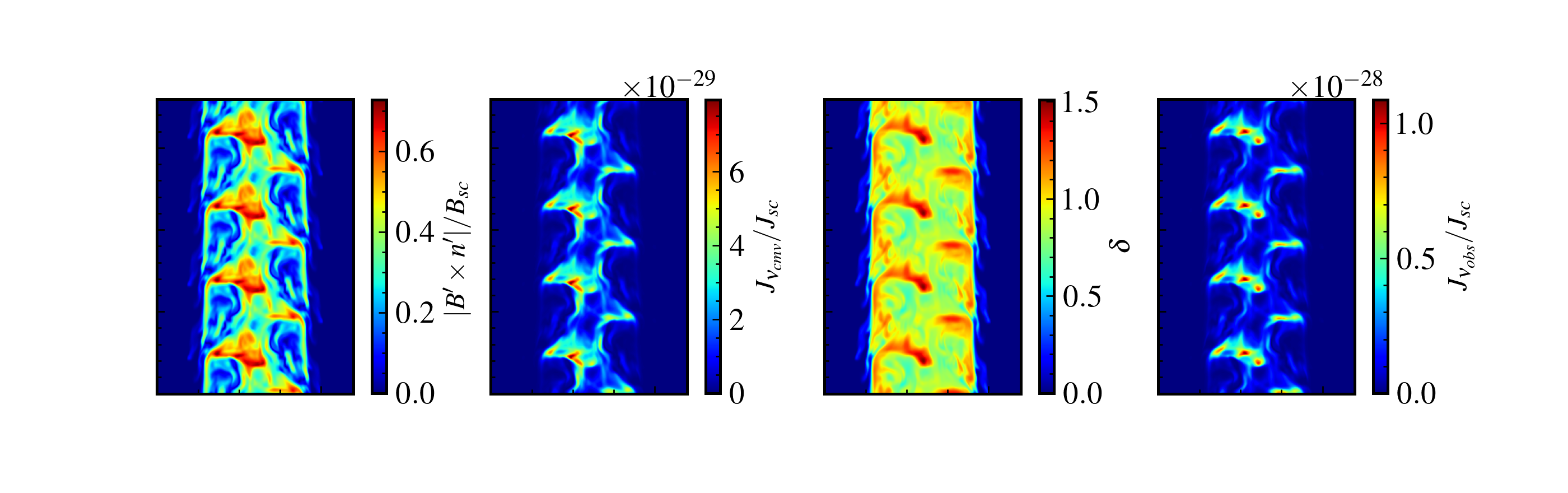}
    \caption{X-Z cuts of $|\bvec{B^{\prime}} \times \bvec{n^{\prime}}|$, synchrotron emissivity in the comoving frame, Doppler Factor, and synchrotron emissivity in the observer frame for the Ref case at $t/t_{\rm sc}$ = 70 for an observer making an angle of 20$^\circ$ (top) and 45$^\circ$ (bottom) with respect to the axis of the column respectively. All the quantities are normalized to their scaled values.}
    \label{fig:Jnu_XZcut}
\end{figure*}

The X-Z cuts of different parameters that may contribute to the emission are shown in figures~\ref{fig:ThEng_XZ} and \ref{fig:Jnu_XZcut}. The total energy density radiated by the chosen particle distribution is considered to be a fraction of the thermal energy density $\varepsilon_{\rm th}$. The X-Z cut of $\varepsilon_{\rm th}$ for the Ref case at $t/t_{\rm sc}$ = 70 is shown in figure~\ref{fig:ThEng_XZ}. The distribution of $\varepsilon_{\rm th}$ is more at the shear boundaries, in high density regions. By using this distribution given in equation~(\ref{eq:thermN}), the normalisation required to determine the non-thermal particle spectra is obtained. Finally, the emissivity in the observer frame can be quantified using the normalised power-law spectral distribution (see equation~(\ref{eq:SyncEms})) which also depends on the angle between the magnetic field vector $\&$ line of sight vector and the Doppler factor.

The emission has contributions from quantities having directional dependence such as the angle between the magnetic field vector and the line of sight vector ($|\bvec{B^{\prime}} \times \bvec{n^{\prime}}|$) and the Doppler factor ($\delta$). Figure~\ref{fig:Jnu_XZcut} represents the X-Z cuts of the quantities such as $|\bvec{B^{\prime}} \times \bvec{n^{\prime}}|$, emissivity in the comoving frame, Doppler factor and the observed emissivity for an observer making an angle of 20$^{\circ}$ (top) and 45$^{\circ}$ (bottom) with the axis of the plasma column respectively for the Ref case at $t/t_{\rm sc}$ = 70. From figures \ref{fig:ThEng_XZ} and \ref{fig:Jnu_XZcut}, it can be seen that both thermal energy density and $|\bvec{B^{\prime}} \times \bvec{n^{\prime}}|$ contribute to the total synchrotron emission in the comoving frame. However, in the observed frame, the contribution from the combination of Doppler factor and $|\bvec{B^{\prime}} \times \bvec{n^{\prime}}|$ is dominant in the synchrotron emission (See section \ref{sec:Int_maps} for more detailed explanation). \cite{Raiteri_2017} suggested that for a jet with helical structure there would be different emitting regions depending on the line of sight of the observer. As seen in the last column of figure \ref{fig:Jnu_XZcut}, the location of high emitting region is different for the observers making angles of 20$^{\circ}$ and 45$^{\circ}$ with respect to the axis of the plasma column. For the observer making 20$^{\circ}$ angle, the value of the Doppler boosting factor is more compared to the scenario where the observer is making 45$^{\circ}$ angle with respect to the axis of the plasma column. As a consequence, the overall emission is more for the observer inclined at 20$^{\circ}$ angle with respect to the axis of the plasma column.

These figures clearly demonstrate the impact of viewing angle and also explain the effect of each contributing factor on the observed emission features.

\subsection{Impact on the observed emission}

\subsubsection{Intensity maps}\label{sec:Int_maps}
\begin{figure*}
    \centering
    \includegraphics[scale=0.43]{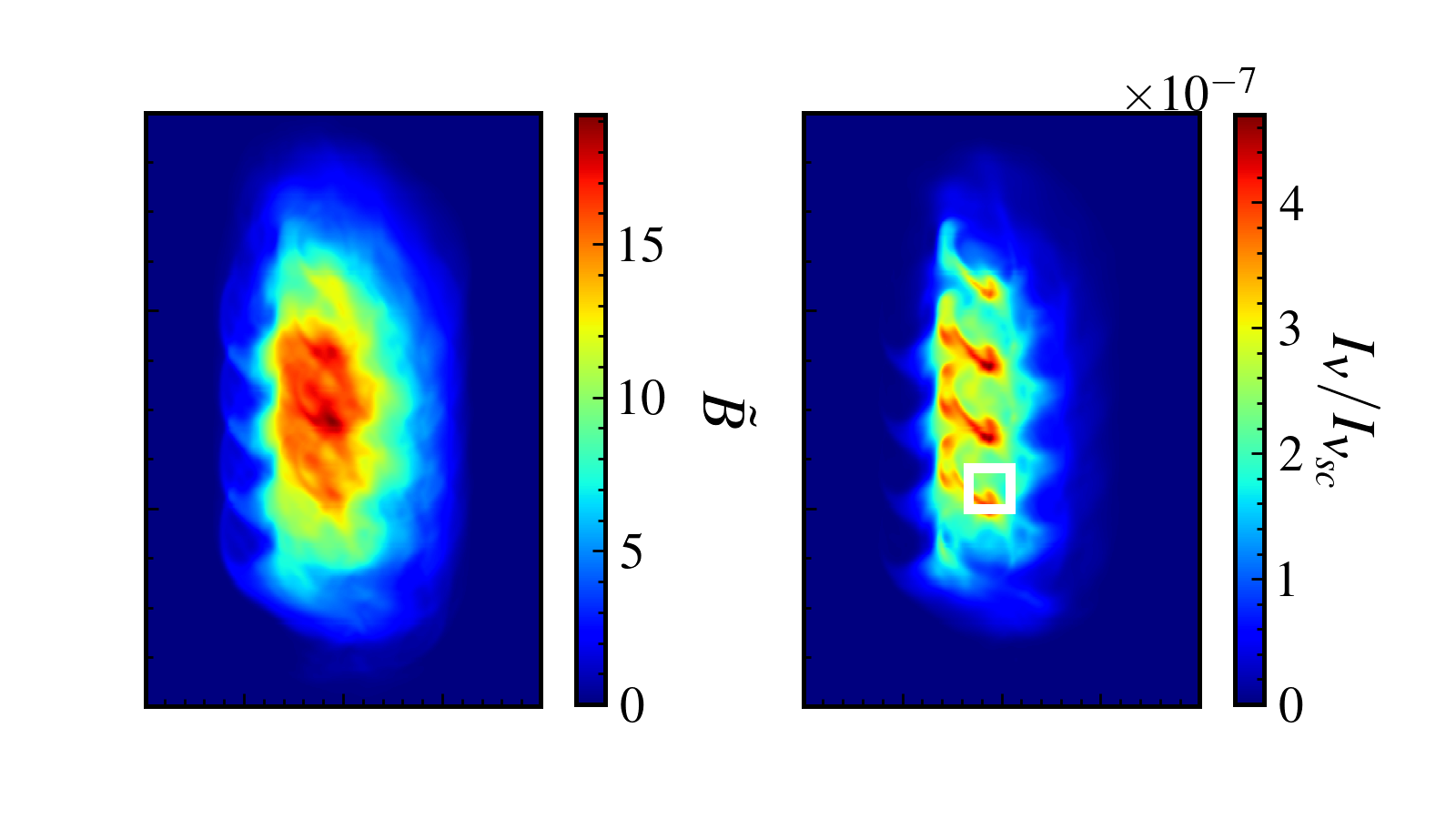}
    \includegraphics[scale=0.43]{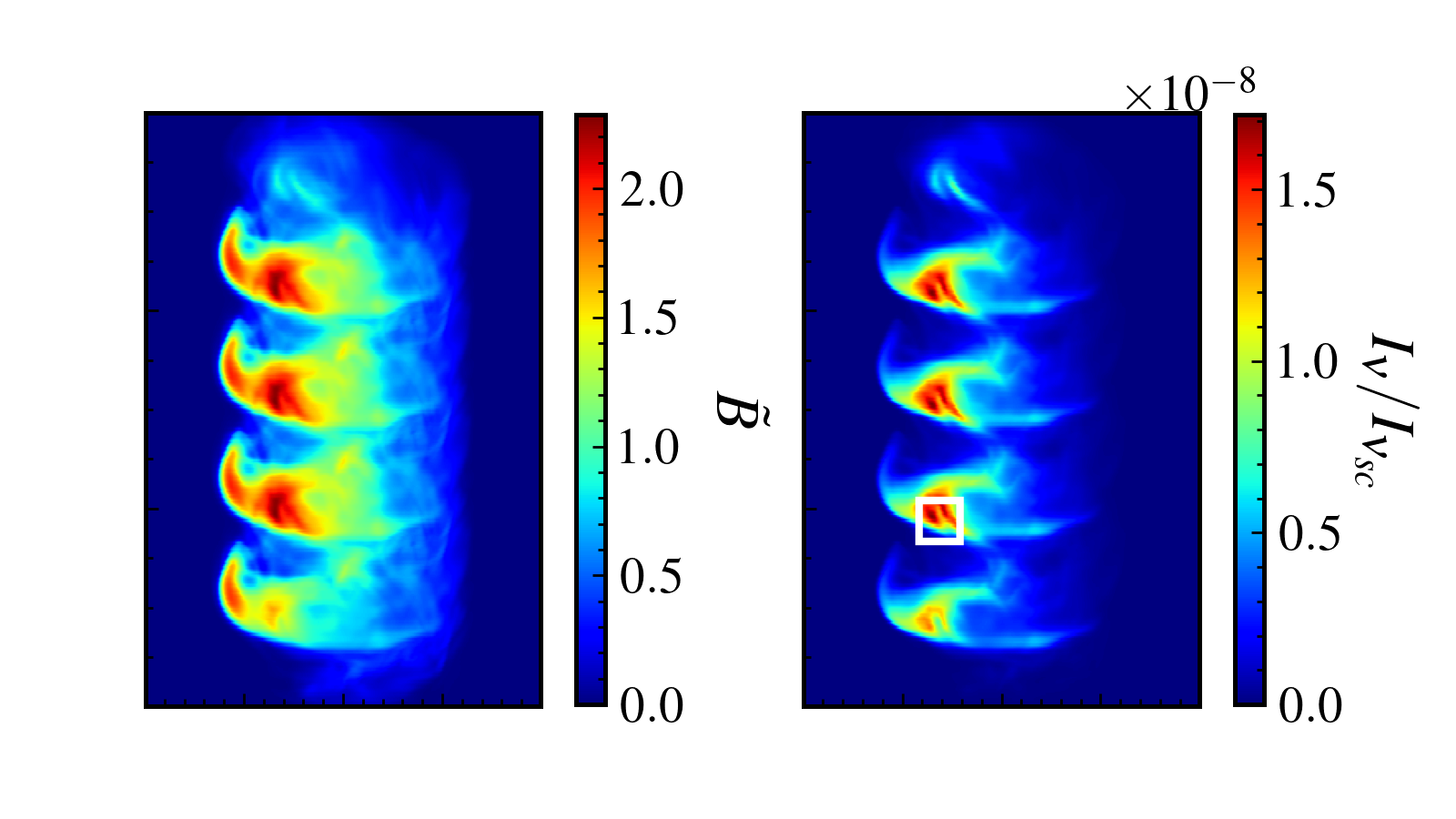}
    \caption{$\tilde{B}$ and $I_{\nu}/I_{{\nu}_{\rm sc}}$ for the Ref case at $t/t_{\rm sc}$ = 70 for an observer making angles of 20$^{\circ}$ and 45$^{\circ}$ respectively with the axis of the plasma column.}
    \label{fig:Bxndel2_Int}
\end{figure*}

The impact of different viewing angles on the emission obtained form the helical jet can also be understood from the intensity maps. The initial conditions for the particle spectra used for estimating the synchrotron intensity are the same as given in section \ref{sec:Sync_plane}. 
Figure \ref{fig:Bxndel2_Int} represents $\tilde{B}$ = $\int |\bvec{B^{\prime}} \times \bvec{n^{\prime}}|\, \delta^{2} ds$ in the units of ($B_{\rm sc} \times L_{\rm sc}$), where $ds$ is the elemental distance along the given line of sight and $I_{\nu}/I_{{\nu}_{\rm sc}}$ for the Ref case at $t/t_{\rm sc}$ = 70 for an observer making angles of 20$^{\circ}$ and 45$^{\circ}$ respectively with the axis of the plasma column. The emission is brighter in the regions of strong $\tilde{B}$. This shows the significance of the viewing angle on the observed emission signatures.

From figure \ref{fig:Ref_dynamics}, we can see that the kink nodes appear to be quasi-periodic in structure. To investigate any periodic nature associated with the kink, we arbitrarily choose a particular section of the plasma column that focuses on just one kink, instead of tracing the whole column. We fix a specific area of the plasma column that covers $25^2$ grid points corresponding to $\approx$ 0.2$L^2_{\rm sc}$. 
We track the evolution of the instability in that particular section, highlighted as the white box shown in figure \ref{fig:Bxndel2_Int}.

\begin{figure*}
    \centering
    \includegraphics[scale=0.5]{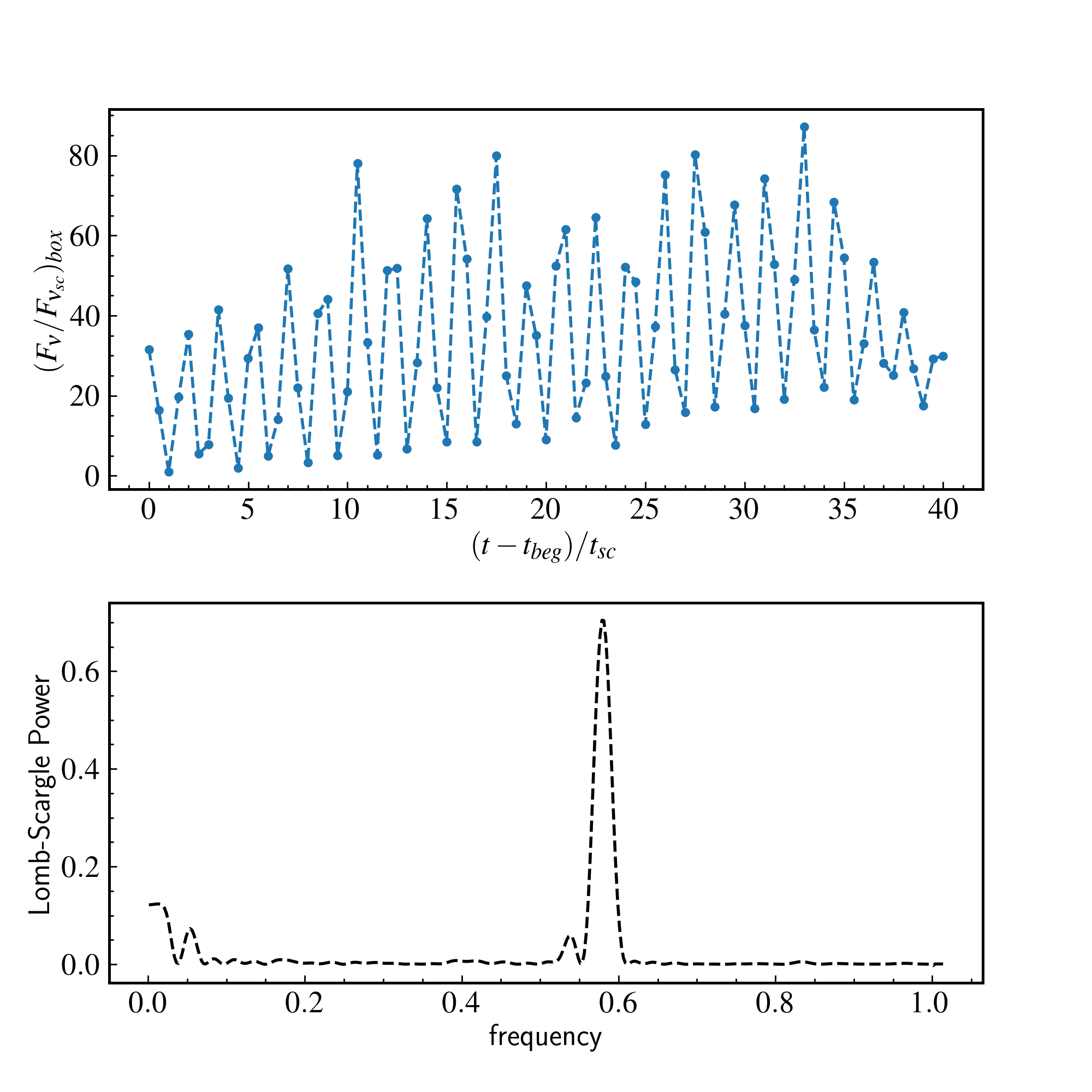}
    \caption{Top panel: Simulated light curve from the highlighted box for the Ref case for an observer making 45$^\circ$ angle with respect to the axis of the plasma column. Here $t_{\rm beg}$ = 40 $t_{\rm sc}$ and $F_{\nu}/F_{\nu_{\rm sc}}$ within the box is normalized to $8.28\times 10^{-8}$. Bottom panel: Lomb-scargle periodogram with periodicity time scale of 1.72 $t_{\rm sc}$.}
    \label{fig:LC_box}
\end{figure*}

The top panel of figure \ref{fig:LC_box} displays the simulated light curve emitted out from the defined box for a time duration of $t/t_{\rm sc}$ = [40-80] for an observer making an angle of 45$^\circ$ with the axis of the plasma column. During this period, the instability has the maximum development and hence, it is expected to see the maximum emission from the plasma column. The curve shows multiple peaks and appears to be quasi-periodic in nature. To analyze the periodicity, we implement the Lomb–Scargle periodogram \citep{VanderPlas_2018_Lomb} and the corresponding power is plotted against the frequency as shown in the bottom panel of figure \ref{fig:LC_box}. The period of the simulated light curve is found to be $t/t_{\rm sc}$ $\approx$ 1.72 with the peak significance value, false alarm probability (FAP) $\ll 1$. We have also calculated the periodicity for a time duration of $t/t_{\rm sc}$ = [40-80] by considering the observer making an angle of 20$^\circ$ with the axis of the plasma column and the time period is found to be $t/t_{\rm sc}$ $\approx$ 1.72 with FAP $\ll 1$ . As shown in figures \ref{fig:Ref_dynamics} and \ref{fig:Ref_4Slices}, the kink not only moves upward but also moves in the transverse direction with the expansion of the plasma column. Hence, we expect the light curve periodicity time-scale to correlate with the dynamical evolution time-scale of the instability. The estimated periodicity is also consistent with the time taken by a single kink to traverse the complete plasma column ($L_{\rm z}$ = 6$L_{\rm sc}$) for the axial speed $v_{\rm z} \sim 0.97c$. The minimum variability time-scale obtained using the formula provided by \cite{Burbidge_1974} is found to be $t/t_{\rm sc} \sim 0.28$. This corresponds to the emitting region of size $\sim 0.3$ in the units of $L_{\rm sc}$ which is consistent with our initial consideration of box size.

\subsubsection{Simulated light curve}\label{sec:simulated_lc}

\begin{figure*}
    \centering
    \includegraphics[scale=0.35]{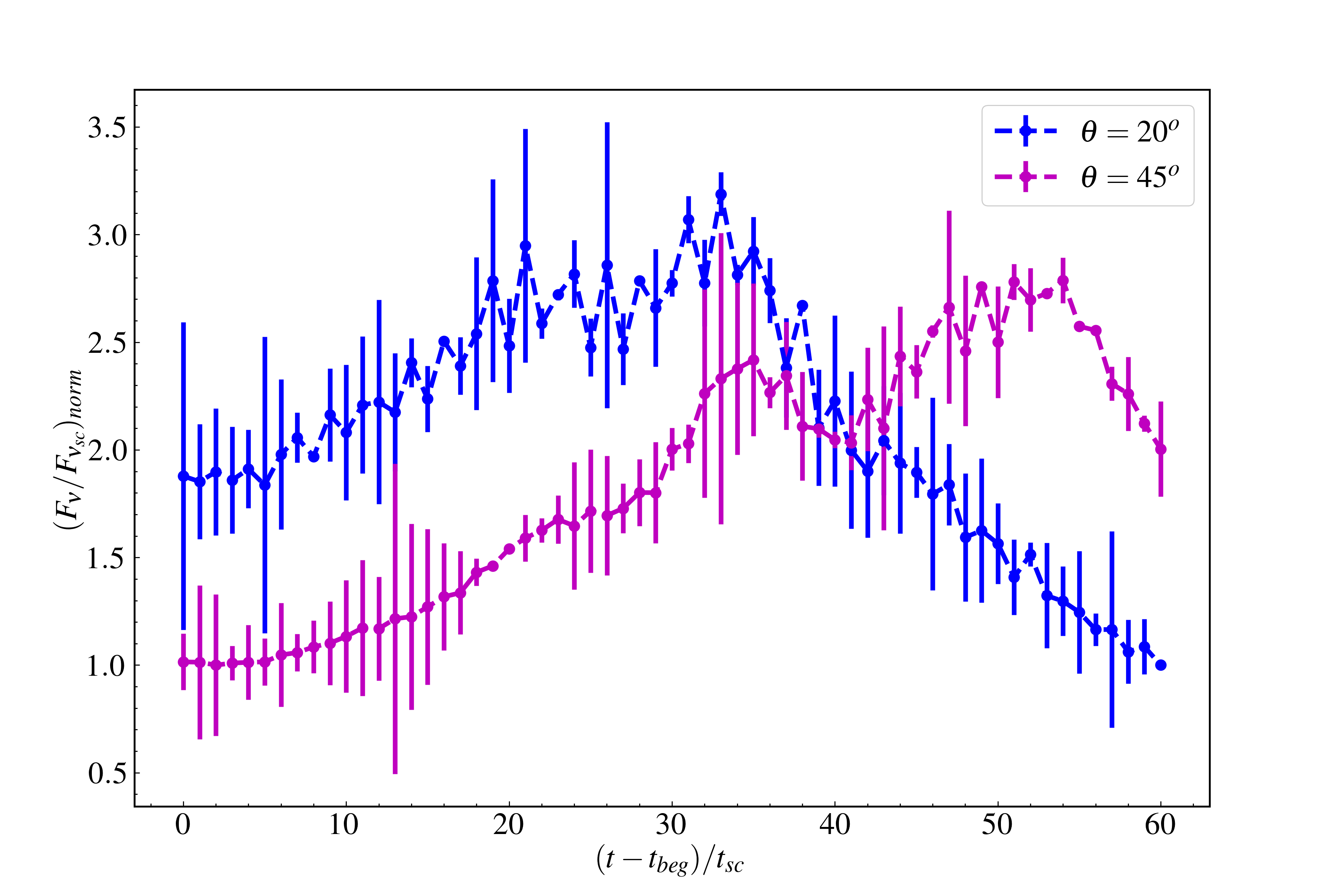}
    \caption{Simulated light curve for the Ref case where $F_{\nu}/F_{\nu_{\rm sc}}$ is normalized to $8.27\times 10^{-24}$ and $2\times 10^{-25}$ for an observer making an angle of 20$^\circ$ and 45$^\circ$ with respect to the axis of the plasma column respectively. Here, $t_{\rm beg}$ = 20 $t_{\rm sc}$ and the vertical lines represent the error bar with SNR = 60.}
    \label{fig:LC_Ref}
\end{figure*}

Another way of understanding the emission features associated with the helical jet configuration is to estimate the total integrated flux.
We have obtained the total integrated flux density for the Ref case with an observer making an angle of 20$^\circ$ and 45$^\circ$ with respect to the axis of the plasma column (see figure \ref{fig:LC_Ref}). The spectral input parameters are the same as given in the previous sections. It should be noted that the simulated light curve shown in figure \ref{fig:LC_box} corresponds to the emission obtained from the box, that covers a single kink. However, the total integrated flux (see figure \ref{fig:LC_Ref}) obtained from the whole plasma column nullifies the periodic nature of kink and provides the composite emission coming from all the kink nodes.

In order to quantify the observed variability, we have performed different statistical tests such as the chi-square statistics ($\chi_{\rm red}$), the fractional root mean square variability ($f_{\rm var}$) and the relative variability amplitude (RVA) on the synthetic light curve (figure~\ref{fig:LC_Ref}).
The details of different variability statistical tests are provided in appendix \ref{app:stats}.

As the synthetic light curve adopted for carrying out the above tests does not have any inherent errors (due to instrumentation etc.), we have also simulated the error bars for each of the simulated flux values. This simulated error is computed for three different SNR values (see section~\ref{emission_setup}). The results of the above mentioned statistics are given in a tabular form in table \ref{tab:stat_test0}. 

\begin{table*}
    \centering
    \caption{Results of the statistical tests for an observer making an angle of 20$^{\circ}$ with respect to the axis of the plasma column.}
    \begin{tabular}{|c|c|c|c|c|c|c|c|c|c|c|}
    \hline\hline
    \multirow{2}{*}{Runs ID} & \multirow{2}{*}{s.d} & \multicolumn{3}{c|}{${\chi^{2}}_{\rm red}$} & \multicolumn{3}{c|}{$f_{\rm var}$} & \multicolumn{3}{c|}{RVA} \\ [2ex]
    \cline{3-11} 
    
      &  & SNR = 20 & SNR = 60 & SNR = $\infty$ & SNR = 20 & SNR = 60 & SNR = $\infty$ & SNR = 20 & SNR = 60 & SNR = $\infty$  \\ [2ex]
    \hline\hline
     
     Ref\_n2 & 2.77  & 7.99$\times 10^{2}$ & 1.31$\times 10^{3}$  & 1.2$\times 10^{11}$ & 0.58$\pm$ 0.029 & 0.6$\pm$ 0.002 & 0.62$\pm 2.9\times 10^{-6}$  & 0.84$\pm$ 0.36 & 0.84$\pm$ 0.26  & 0.84$\pm$ 0.000034 \\ [2ex]
     
     Ref\_n3 & 1.84 & 1.35$\times 10^{2}$ & 1.72$\times 10^{2}$ & 1.3$\times 10^{10}$ & 0.45$\pm$ 0.032 & 0.49$\pm$ 0.017 & 0.51$\pm$ 3.2$\times 10^{-6}$ & 0.75$\pm$ 0.64 & 0.75$\pm$0.21 & 0.75$\pm$ 0.000048 \\ [2ex]
     
     Ref & 0.55 & 1.33$\times 10^{1}$ & 1.87$\times 10^{2}$  & 1.35$\times 10^{9}$  & 0.11$\pm$ 0.058  & 0.23$\pm$ 0.01 & 0.26$\pm 2.8\times 10^{-6}$ &  0.52$\pm$ 0.41 & 0.52$\pm$ 0.19  & 0.52$\pm$ 0.000043 \\ [2ex]
     
     Ref\_n8 & 0.14  & 6.2  & 23.01 & 4.2$\times 10^{8}$ & --  & -- & 0.11$\pm$ 2.9$\times 10^{-6}$ & 0.18$\pm$ -- & 0.18$\pm$ --  & 0.18$\pm$ 0.000036 \\ [2ex]
     \hline
    \end{tabular}
    Column 1: runs ID; column 2: standard deviation of the data set; column 3: reduced $\chi^{2}$ value; column 4: root mean square fractional variability; column 5: relative variability amplitude.
    \label{tab:stat_test0}
\end{table*}

The ${\chi^{2}}_{\rm red}$ value is maximum for Ref\_n2 compared to the runs Ref\_n3, Ref, and Ref\_n8 for all three SNR values. This implies that the case with $n=2$ has the maximum variablity with respect to its mean value. RVA is also found to be maximum with a value of $0.84\pm(0.36,0.26,0.000034)$ for Ref\_n2, where the errors are obtained for increasing values of SNR.  
Also, the run Ref\_n2 has the maximum $f_{\rm var}$ value of $0.58\pm 0.016$ with SNR = 20. Whereas, for Ref\_n8, the kink growth is suppressed and hence the variability is not significant. The random errors added to the simulated data, overpowered the variability and we did not obtain any $f_{\rm var}$ values with lower SNR values. $f_{\rm var}$ values represent a stronger variability in Ref\_n2 and Ref\_n3 cases in comparison with Ref and Ref\_n8 cases. We have also performed the statistical tests for the Ref case with an observer making angles of 5$^{\circ}$ and 45$^{\circ}$ with respect to the plasma column axis. We obtained stronger variability as the observer moved closer to the axis of the plasma column. The RVA is found to be 0.77$\pm$ 0.29, 0.52$\pm$ 0.19 and 0.47$\pm$ 0.22 for the viewing angles 5$^{\circ}$, 20$^{\circ}$ and 45$^{\circ}$ respectively for the Ref case with a moderate SNR value.

It should be noted that the analysis is performed for a section of the simulated flux data spanning nearly 20-30 years in physical units. 
In summary, all the above statistical analysis indicates the existence of long-term variability linked with the helical jet structure.

\section{Discussion}\label{discussion}

In this work, we have carried out a high resolution 3D relativistic MHD simulation of a plasma column with seven different initial conditions and parameters. Three different parameters such as $n$, $\sigma$ and $\Gamma$ have been chosen and the impact of these parameters on the dynamical and emission properties have been studied. Among the runs Ref\_n2, Ref\_n3, Ref and Ref\_n8, which differ in $n$ values with all the other physical parameters being the same, Ref\_n2 exhibits the maximum kink growth rate. In our work, the growth rate has been calculated as a consequence of the evolution of the transverse kinetic energy. \cite{Appl_2000} have performed the linear theory analysis of the $m=1$ mode instability and determined the growth rate of the fastest growing mode. Our result is in agreement with their analysis. We also estimated the magnetic energy dissipation rate from the non-linear phase of the instability. We found a similar trend of dissipation rate as that of growth rate for different runs. More is the growth of the perturbation, the system would be more turbulent, and consequently, the dissipation rate would be higher. In section \ref{results}, we discuss the effects of the growth rate on the morphological structure of the plasma column and we see that the deviation of the density barycentre from the axis of the plasma column is maximum for the case with the maximum growth rate. Furthermore, high Lorentz factor ($\Gamma$ = 10) weakens the formation of the kink and that gets reflected in the energetics and the density barycentre deviation. Additionally, we study the effect of changing the magnetization on the evolution of the kink instability. In the $\sigma$ = 1 runs, the plasma matter is mostly distributed near the axis of the column, \EC{indicating a slower} development of the kink. \EC{In the case with low magnetization value along with the presence of shear, mixing of both kink and KH instabilities is expected due to trans-Alfv\'enic nature of the flow.}
    
\begin{figure}
        \centering
        \includegraphics[scale=0.2]{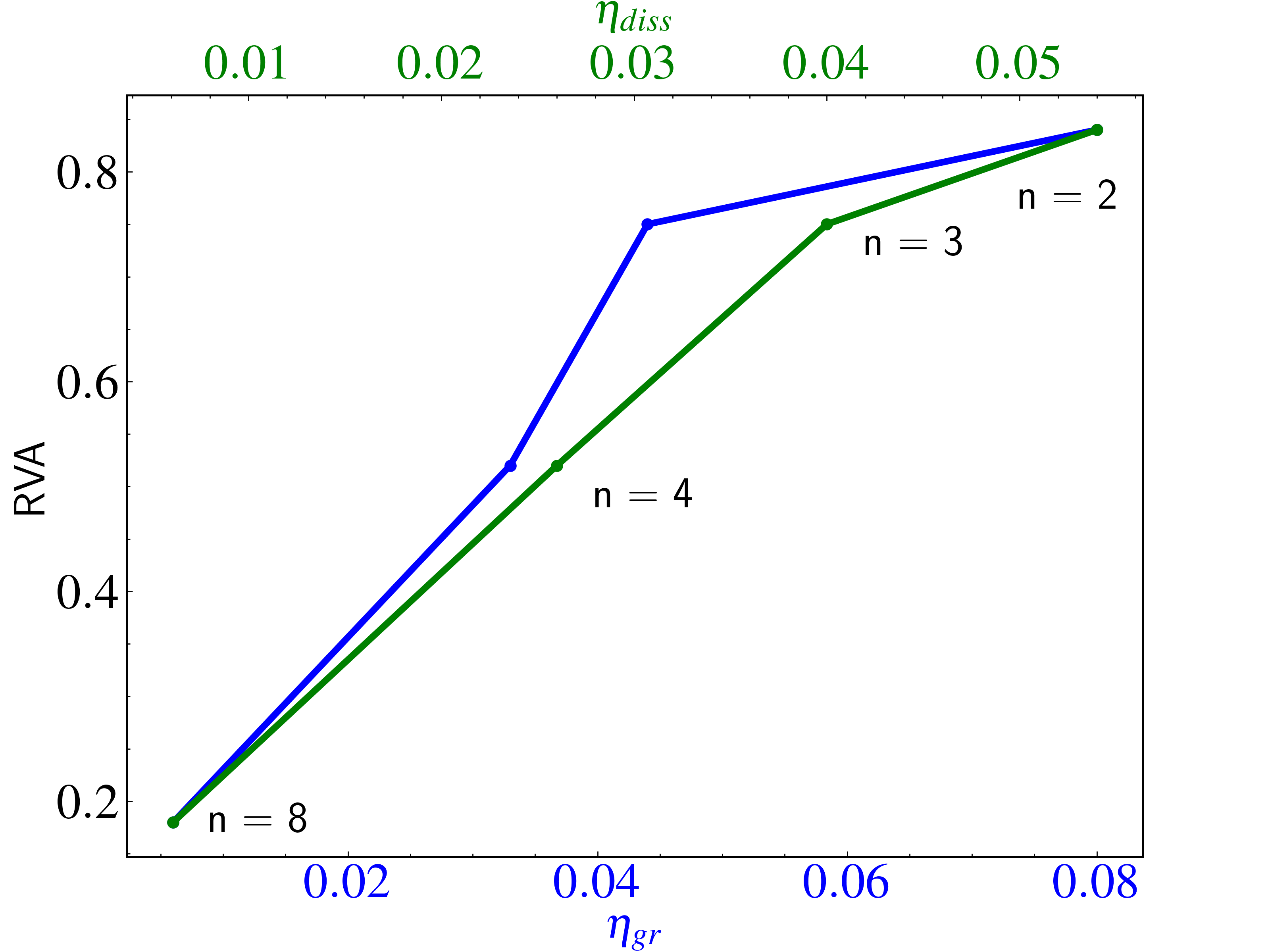}
        \caption{Plot of kink growth rate ($\eta_{\rm gr}$) $\&$ dissipation rate ($\eta_{\rm diss}$) and RVA for the runs Ref\_n2, Ref\_n3, Ref $\&$ Ref\_n8.}
        \label{fig:RVAGrate}
    \end{figure}

The kink growth rate is one of the important parameters that describes the evolution of the instability. \cite{Dong_2020} showed that the kink instability in blazar jets may cause quasi-periodic oscillations (QPO) and the period of QPOs is associated with the growth time of the kink instability. Their simulation is characterized with a modest Lorentz factor $\Gamma \approx$ 2, which is less than the typical value of the Lorentz factor relevant in the case of Blazar jets. Higher kink growth rate would result in a greater distortion of the plasma column. In our work, we have considered a moderately relativistic scenario with $\Gamma$ = 5 and 10, and we correlate the kink growth rate with the variability detected from the simulated light curve. Relative variability amplitude (RVA) calculated in section \ref{sec:simulated_lc}, quantifies the difference existing between the maximum and minimum value of the simulated flux density, irrespective of the intrinsic error. Figure \ref{fig:RVAGrate} shows a correlated trend of RVA with the kink growth rate ($\eta_{\rm gr}$) and dissipation rate ($\eta_{\rm diss}$). We find that the RVA is maximum for the case with the lowest $n$ value, having the maximum kink growth rate. A high RVA value indicates a stronger kinked jet with high magnetization and a moderate Lorentz factor. The correlated trend between RVA and $\eta_{\rm gr}$ implies that highly magnetized jets being subjected to the kink instability would have a high value of RVA, thus exhibiting strong flux variability. A faster dissipation rate of the volume averaged transverse magnetic energy in the non-linear growth regime corresponds to a faster decaying of the light curve. Further, among the runs with $\Gamma$ = 5, all the statistical analysis (${\chi^{2}}_{\rm red}$, $f_{\rm var}$, RVA) indicates that the variability is maximum for the case with the highest kink growth rate and dissipation rate. 
    
The effect of viewing angle in the context of intensity maps has been studied in section~\ref{sec:Int_maps}. Besides, the impact of viewing angle on the simulated light curve is also studied. The plasma column moves in the Z-direction as the axial component of the velocity vector is the dominant one. When the observer moves closer to the axis, the emitting region of the plasma column makes a smaller angle and hence provides a higher Doppler boosting factor. A high $\delta$ value gives a higher amplitude of simulated flux, consequently a stronger variability. The RVA, which particularly measures the variability amplitude that exists in the simulated data, increases for the Ref case as the observer's line of sight makes a smaller angle with respect to the axis of the column. For an observer making a 5$^{\circ}$ angle, the Doppler factor would be higher compared to an observer making 20$^{\circ}$ and 45$^{\circ}$ angle. As a result, the RVA increases as the observer becomes more inclined towards the axis of the column.    
    
The observed helical structure of the jet may have several origins, such as the presence of a binary black hole system, the precession or rotation of the jet or due to the presence of MHD instabilities. The geometry of the jet causes different jet regions to change their orientation and hence, their relative Doppler factors. \cite{Villata_1998} has used this model to explain the quasi-periodic double peaked structure of the optical outbursts observed in the blazar OJ 287. In this scenario, a double jet emerges out of two black holes in a binary system and that explains 
each peak in the double-peaked structure is due to one of the jets, where the changing intensity of the light curve is due to the non-alignment of the orbital axis with the line of sight of the observer. Further, a variable Doppler factor as a result of the helical structure can well explain the observed long-term behaviour of many blazars jets (for example see \cite{Villata_1999, Ostorero_2004, Raiteri_2017}). Recently, a geometric explanation for the short-term flux variation time-scale is given by \cite{Raiteri_2021}, where they consider a twisted jet, composed of many filaments where each filament has its own orientation with respect to the observer and thus its own changing Doppler factor.
    
To quantify the variability, we have performed different statistical tests by incorporating the standard error of the values obtained for the simulated light curve. 
\cite{Schleicher_2019} have analyzed the variability of two bright blazars Mrk 421 and Mrk 501 in various energy ranges using the fractional variability ($f_{\rm var}$). The values of $f_{\rm var}$, we obtained in our study are within the range of values obtained by \cite{Schleicher_2019} in the optical band with a viewing angle $\approx$ 3-5 times smaller compared to the typical value considered in Blazar jets. 
However, a much detailed and systematic study is required to understand the dependence of the fractional variability and the energy. 
At the same time, it should also be noted that we simulate a particular portion of the jet, not the whole jet from its launching to termination. We focus on a smaller section of the jet, that could capture the local features such as instabilities and bear the computational expense with high grid resolution. This chosen resolution of 30 grid cells per jet radius is sufficiently high to capture the instability growth and helps in generating the associated synthetic light curve. Therefore, it is expected that similar emission signatures are likely to occur in other regions of the jet.
Our emission modeling approach accounts for the changes in the physical properties such as the fluid density, the bulk flow velocity, and the magnetic field strength but does not incorporate the evolution of the emitting particle spectra.
    
\section{Summary}\label{summary}
Blazars are a sub-class of AGN jets that show multi-timescale variable signatures. Additionally, they also show the presence of flaring in multiple wavebands. The geometric model that requires the presence of a twisted jet has been proposed to explain the long term variability \cite{Raiteri_2017}. The present work is motivated by this model and aims to relate dynamical features due to the relativistic MHD kink instability with the variability signatures. 
For this purpose, we have carried out high resolution 3D RMHD simulations with varying parameters of a plasma column as a representative section of the parsec scale jet.

Our main results can be broadly classified as follows: 
\begin{enumerate}
  
 \item \textbf{Dynamical impact of kink growth on Twisted Jet Model -}
In this work, we have studied the dependence of the crucial parameters on the growth of the kink instability through a slew of parameter runs. 
\EC{Among the numerical setups studied in this work, we have obtained the maximum kink growth rate for an axial wavenumber $n$ = 2. However, from linear theory analysis, with our considered initial conditions, $n$ = 1 shows the maximum growth rate, and its value is expected to get reduced with an increasing value of $n$. This is consistent with the results obtained from our simulations. Further, we observe a stalled growth of kink mode with a lower value of the magnetization parameter $\sigma$ as the flow does not remain  sub-Alfv\'enic during the considered evolution time.}
Additionally, the onset of the kink instability is delayed, and also the growth of the kink mode is reduced for a higher value of the Lorentz factor $\Gamma$. 
We have also demonstrated the effect of the growth of the instability by defining the density barycentre. For the cases with different axial wave-numbers ($n$), the displacement of the column from its axis is more for the case with the higher growth rate. Similarly, the barycentre displacement is higher for the Ref case than the Ref\_g10 case. (See section~\ref{sec:parameter_study}). The consequence of the development of the instability is also contemplated in the behaviour of the averaged energies (see figure \ref{fig:energy_g5}).

 \item \textbf{Role of kink growth in variability signatures -}
We have carried out various statistical estimates from the synthetic light curves obtained from our parametric study including errors generated from the user-defined SNR. 
These estimates connect the dynamics with the emission features and support the helical jet picture as a model to explain the long term flux variation for a period of $\leq$ 20 years. At first, we focus on a single kink and the emission associated with it is found to be periodic where the periodicity time scale is consistent with the dynamical time taken by a single kink to traverse the whole plasma column. We also obtained the minimum variability time scale, that typically provides the information about the size of the emitting region. 
For the reference case with the line of sight at 20$^{\circ}$ with respect to the axis of the jet, we obtained a strong variability in the R-band for modest SNR values.

The parameter study carried out as a part of this work has also allowed us to formulate a relation between the kink growth rate (measured using the kinetic energy evolution) and magnetic energy dissipation rate (measured using transverse magnetic energy) with Relative Variability Amplitude (measured from the synthetic light curve). The large variation observed in the synthetic light curve can be attributed to the growth of the instability as well as the faster dissipation rate of the magnetic energy. The empirical trend obtained from our synthetic light curves shown in figure ~\ref{fig:RVAGrate} provides a one to one co-relation between the linear growth rate and the magnetic energy dissipation rate with RVA which is an observable quantity. 
\end{enumerate}

One of the limitations of our emission modelling method is that the particle distribution does not evolve with time. 
In the subsequent work, we aim to extend the simulations adopting a non-thermal emission modeling method with evolving particle spectra using the hybrid macro-particle based framework in the PLUTO code developed by \cite{Vaidya_2018}. The synthetic light curves obtained using such a hybrid model would provide additional multi-waveband information for a better understanding of the physical processes responsible for the high energy and variable emission of AGN jets.

\section*{Acknowledgements}
\EC{The authors would like to thank the referee for the constructive comments and suggestions that helped to improve the manuscript signiﬁcantly.
The authors would also like to thank Gianluigi Bodo and Yosuke Mizuno for their valuable comments and useful suggestions on the manuscript. SA is supported by the DST INSPIRE Fellowship and would like to acknowledge the support for Ph.D. BV would like to acknowledge the support from the Max Planck Partner Group Award. All computations presented in this work are carried out using the facilities provided at IIT Indore and the Max Planck Institute for Astronomy Cluster: ISAAC which is a part of the Max Planck Computing and Data Facility (MPCDF).} 

\section*{Data Availability Statement}
The data obtained from this work will be available with considerable request to the corresponding author(s).

\bibliographystyle{mnras}
\bibliography{reference}

\appendix

\section{Effect of initial and boundary conditions}\label{app:effect_bc}

\begin{figure*}
    \centering
    \includegraphics[scale=0.4]{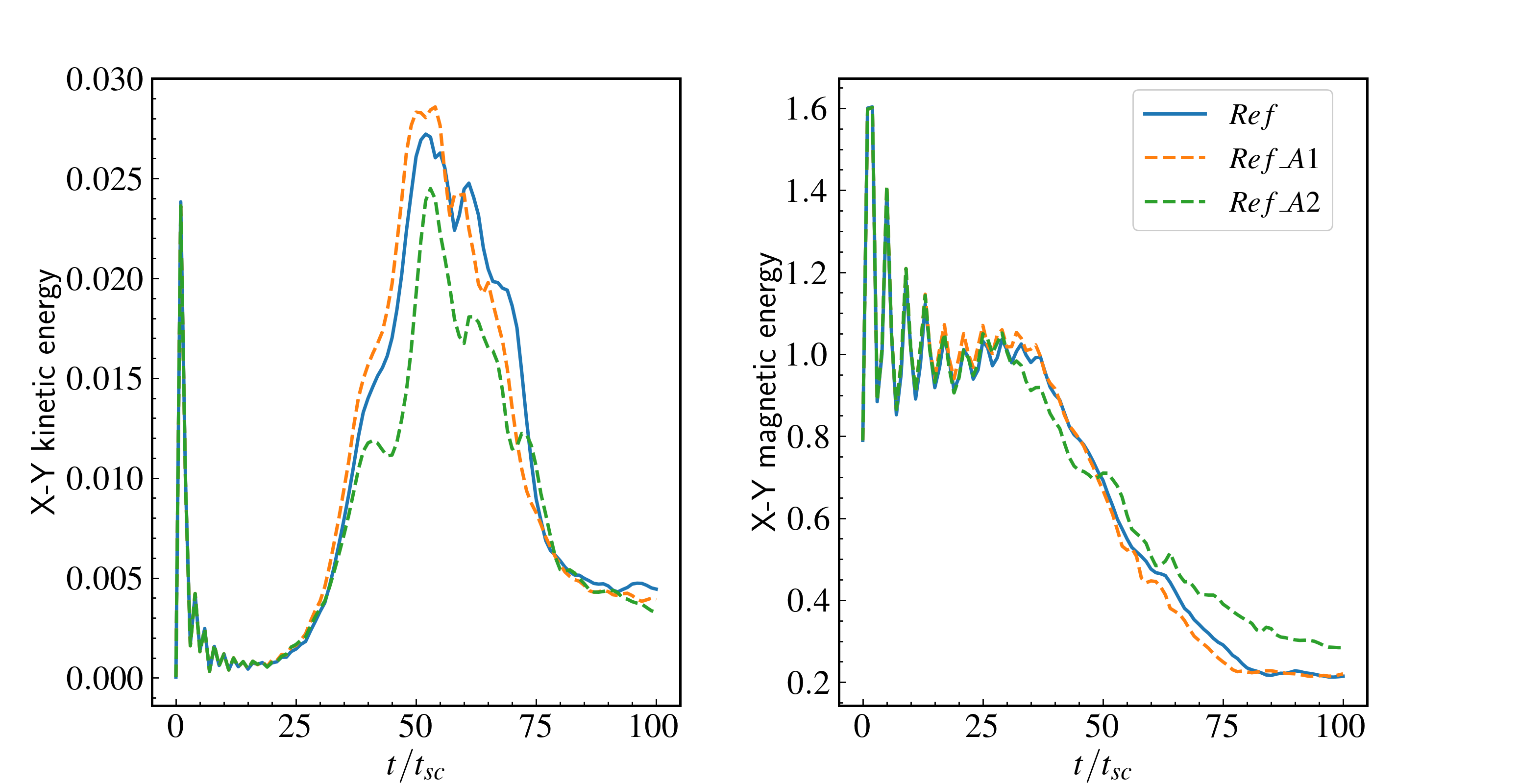}
    \caption{Time evolution of the volume averaged quantities for the cases Ref, Ref\_A1 and Ref\_A2.}
    \label{fig:app_energy}
\end{figure*}

We have investigated the effect of different initial and boundary conditions on the growth of the instability. Since a radially dependent pressure profile does not favor the system to be in equilibrium initially, we have performed a simulation for our Ref case with constant initial pressure, named as Ref\_A1. 
We have also studied the effect of boundary conditions by fixing the transverse simulation box boundary to be reflective with constant initial pressure, named Ref\_A2. We have shown the evolution of transverse kinetic and magnetic energy for the cases Ref\_A1 and Ref\_A2 in comparison with the Ref case in figure \ref{fig:app_energy}. We do not observe any notable difference in both curves. The onset of the instability is the same for all three cases with inappreciable variation in the non-linear regime. The dissipation of transverse magnetic energy is approximately the same, indicating a similar development and growth of the instability. We also observe emission signatures associated with these configurations are similar to the results provided in section \ref{helical_jet}.

\subsection{Discussion on equilibrium}\label{sec:equi_app}

\begin{figure*}
    \centering
    \includegraphics[scale=0.62]{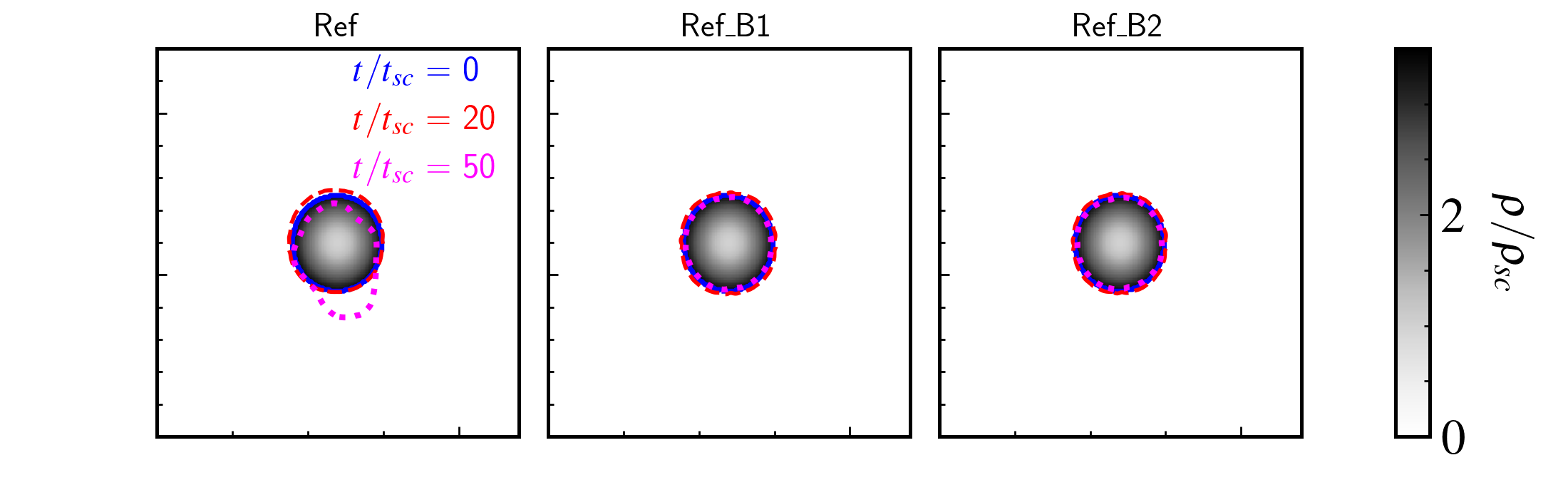}
    \caption{\EC{X-Y distribution of plasma column density at Z = 0.5$L_{\rm z}$ at $t/t_{\rm sc}$ = 0 for all the 3 cases. Over-plotted as blue solid, red dashed and magenta dotted lines are the tracer contours of level = 0.9 at $t/t_{\rm sc}$ = 0, 20 and 50 respectively.}}
    \label{fig:rhoTr_contour}
\end{figure*}

\EC{To understand the effect of an electric field on the radial equilibrium, we study the evolution of the plasma column for a case with a similar configuration as that of the Ref case without perturbation, named as Ref\_B1. Similarly, we also investigate the evolution of a case with a constant pressure profile and with no perturbation, named as Ref\_B2. 
The X-Y distribution of the plasma column density at Z = 0.5$L_{\rm z}$ at $t/t_{\rm sc}$ = 0 for all the 3 cases is shown in figure \ref{fig:rhoTr_contour}. To mark the boundary between the ambient and the column, the tracer contours at level = 0.9 are plotted as blue solid, red dashed and magenta dotted lines at $t/t_{\rm sc}$ = 0, 20, and 50 respectively. At $t/t_{\rm sc}$ = 50, there is enough growth of the instability to take a helical structure in the Ref case; the column width has not changed in the Ref\_B1 and Ref\_B2 cases. Due to loss of equilibrium, a velocity in the radial direction is generated. However, at most, the radial velocity is $\approx$ 1-2\% of the axial velocity during the simulation time considered in our work in both cases without perturbation. This suggests that the radial balance does not affect the structure of the plasma column. Hence, it is expected that the qualitative nature of the growth rate of the instability and the associated emission features will remain unaffected.}

\section{Formulation of variability tests}\label{app:stats}

The chi-square test quantifies the difference between the observed data and the model data. It is defined as \citep{stat_method_2013}: 
\begin{equation}
\chi^{2} = \sum_{i=1}^{N} \frac{(F_{\rm i}-F_{\rm model})^{2}}{{\sigma_{\rm i}}^{2}}.
\end{equation}

For the present analysis, $F_{\rm model} = \bar{F}$ is the weighted mean of the observed flux and $F_{\rm i}$ is the observed flux at the $i$th data point with error $\sigma_{\rm i}$. 
We calculated the reduced ${\chi^{2}}_{\rm red} = \frac{\chi^{2}}{N-1}$, where $N$ is number of measurements with $N-1$ degrees of freedom. 
In general, if the ${\chi^{2}}_{\rm red}$ value is nearly equal to $1$, the model is well fitted to the observed data. 
However, in our case, higher value of ${\chi^{2}}_{\rm red}$ implies higher variation about the mean value while, and smaller ${\chi^{2}}_{\rm red}$ value would mean minimal variation in the flux as compared to the mean flux state. 

Further, we estimate the fractional variability amplitude $f_{\rm var}$ \citep{vaughan_2003, f_var_2016} which gives information about the intrinsic variability amplitude of the source relative to the mean count overcoming the measurement error for each data point. 
It is defined as : 
\begin{equation}
f_{\rm var} = \sqrt{\frac{S^{2}-{\bar{\sigma}}^{2}_{\rm err}}{{\bar{F}}^{2}}}. 
\end{equation}
The error in $f_{\rm var}$ can also be estimated as follows:
\begin{equation}
err(f_{\rm var}) = \sqrt{\left(\sqrt{\frac{1}{2N}}\frac{{\bar{\sigma}}^{2}_{\rm err}}{\bar{F}^{2}f_{\rm var}}\right)^{2} + \left(\sqrt{\frac{{\bar{\sigma}}^{2}_{\rm err}}{N}}\frac{1}{\bar{F}}\right)^{2}},
\end{equation}
where, $S^{2}$ is the sample variance, and ${\bar{\sigma}}^{2}_{\rm err}$ is the mean square error.

Another integral parameter that can quantify the measure of variablity from light curves is the relative variability amplitude (RVA) or the variability index \citep{Kovalev_2005, KSingh_2019}, defined as:
\begin{equation}
    RVA = \frac{F_{\rm max} - F_{\rm min}}{F_{\rm max} + F_{\rm min}},
\end{equation}
and the uncertainty on RVA is given by:
\begin{equation}
    \Delta RV A = \frac{2}{(F_{\rm max}+F_{\rm min})^{2}}\sqrt{(F_{\rm max}\Delta F_{\rm min})^{2} + (F_{\rm min}\Delta F_{\rm max})^{2}},
\end{equation}
where $F_{\rm max}$ and $F_{\rm min}$ are the maximum and minimum values of the simulated flux with $\Delta F_{\rm max}$ and $\Delta F_{\rm min}$ uncertainties respectively.

\bsp	
\label{lastpage}
\end{document}